\documentclass{aastex}
\usepackage{emulateapj5}
\usepackage{apjfonts}
\usepackage{epsf}

\slugcomment{UTAP}
\shorttitle{SMALL SCALE CLUSTERING IN ISOTROPIC ARRIVAL DISTRIBUTION OF UHECRS}
\shortauthors{YOSHIGUCHI ET AL.}
\begin{document}
\title{Small Scale Clustering in the Isotropic Arrival Distribution of
Ultra-High Energy Cosmic Rays
and Implications for Their Source Candidates}
%
\author{Hiroyuki Yoshiguchi\altaffilmark{1}, Shigehiro Nagataki\altaffilmark{1}, Sinya Tsubaki\altaffilmark{1}
 and Katsuhiko Sato\altaffilmark{1,2}}

\altaffiltext{1}{Department of Physics, School of Science, the University
of Tokyo, 7-3-1 Hongo, Bunkyoku, Tokyo 113-0033, Japan}
\altaffiltext{2}{Research Center for the Early Universe, School of
Science, the University of Tokyo,
7-3-1 Hongo, Bunkyoku, Tokyo 113-0033, Japan}


\email{hiroyuki@utap.phys.s.u-tokyo.ac.jp}
%
\received{}
\accepted{}
\begin{abstract}
We present numerical simulations on the propagation of UHE
protons with energies of $(10^{19.5}-10^{22})$ eV
in extragalactic magnetic fields over 1 Gpc.
We use the ORS galaxy sample, which allow us to accurately
quantify the contribution of nearby sources to the energy
spectrum and the arrival distribution, as a source model.
The sample is corrected taking the selection effect and absence
of galaxies in the zone of avoidance ($|b|<20^{\circ}$) into account.
We calculate three observable quantities, cosmic ray spectrum,
harmonic amplitude, and two point correlation function
from our data of numerical simulations.
With these quantities, we compare the results of our numerical calculations
with the observation.
We find that the arrival distribution of UHECRs become to be most isotropic
as restricting sources to luminous galaxies $(M_{\rm {lim}}=-20.5)$.
However, it is not isotropic enough to be consistent with the
AGASA observation, even for $M_{\rm {lim}}=-20.5$.
In order to obtain sufficiently isotropic arrival distribution,
we randomly select sources, which contribute to the observed cosmic ray flux,
from the ORS sample more luminous than $-20.5$ mag,
and investigate dependence of the results on their number.
We show that the three observable quantities including the GZK
cutoff of the energy spectrum can be reproduced
in the case that the number fraction $\sim 10^{-1.7}$ of the ORS
galaxies more luminous than $-20.5$ mag is selected as UHECR sources.
In terms of the source number density, this constraint
corresponds to $\sim 10^{-6}$ Mpc$^{-3}$.
However, since mean number of sources within the GZK sphere
is only $\sim 0.5$ in this case,
the AGASA 8 events above $10^{20.0}$ eV, which do not constitute
the clustered events with each other, can not be reproduced.
On the other hand, if the cosmic ray flux measured by the HiRes,
which is consistent with the GZK cutoff, is correct and
observational features about the arrival distribution of UHECRs are same as
the AGASA, our source model can explain both the arrival distribution
and the flux at the same time.
Thus, we conclude that large fraction of the AGASA 8 events
above $10^{20}$ eV might
originate in the topdown scenarios, or that the cosmic ray flux
measured by the HiRes experiment might be better.
We also discuss the origin of UHECRs below $10^{20.0}$ eV
through comparisons between the number density of astrophysical
source candidates and our result ($\sim 10^{-6}$ Mpc$^{-3}$).
\end{abstract} 
%
\section{INTRODUCTION} \label{intro}
In spite of a increasing amount of data, the origin of cosmic rays
in particular at the highest energy ($\sim 10^{20}$eV) is still unknown.
It is very difficult to produce such ultra-high energy cosmic rays
(hereafter UHECRs) in astrophysical shocks,
which are thought to be responsible for the galactic cosmic rays.
The gyroradii of UHECRs with energies of 10$^{20}$eV are
larger than our Galaxy, and this suggests an extragalactic origin
for any astrophysical scenario.
Using a Hillas-plot \citep*{hillas84,selvon00},
active galactic nuclei (AGN), gamma-ray bursts (GRBs),
dead quasars and colliding galaxies are considered as probable
candidates.
These conventional acceleration models are called bottom-up scenarios.
On the other hand, this situation has in recent years triggered
a number of exotic production models based
on new physics beyond the Standard model of the particle physics
\citep*[see ][ and references therein]{bhattacharjee00},
which are called top-down scenarios.
In such scenarios, UHECRs owe their origin to decay of
some supermassive particles, which could be produced from
Topological Defects (TDs), or be certain metastable supermassive
relic particles (MSRPs) produced in the early universe.

In bottom-up scenarios in which protons are accelerated in
the potential UHECRs sources, like GRBs and AGNs, the large distances
to the earth lead to another problem, called the GZK effect
\citep*{greisen66,zatsepin66}.
UHE protons with energy above 8$\times 10^{19}$ eV interact with photons of
Cosmic Microwave Background (CMB) and lose their energy rapidly due to
the production of pions over distances of tens of Mpc \citep*{stanev00}.
This should results in a cut-off of the cosmic ray spectrum
at energy around 8$\times 10^{19}$ eV.
The situation for nuclei is considered to be worse due to
the photo-disintegration mechanism \citep*{stanev00}.
Observationally, there seems to be a disagreement between the AGASA
which revealed the extension of the cosmic ray energy spectrum
above $10^{20}$ eV \citep*{takeda98} and the High Resolution Fly's Eye
\citep*[HiRes; ][]{wilkinson99} which recently reported the cosmic ray
flux with the GZK cut-off \citep*{abu02}.

Since UHECRs above $10^{20}$ eV observed by the AGASA
must originate within the GZK
sphere, whose radius is typically about 50 Mpc,
the arrival directions are expected to point toward their sources
within a few degrees, which is typical deflection angle of UHECRs
\citep*[e.g.,][]{blasi98}.
However, no plausible source counterparts have been found within the
GZK sphere.
There is also a problem concerning the arrival directions of UHECRs
\citep*[e.g.,][]{blasi98}.
Their distribution seems to be isotropic on a large scale
with a statistically significant small scale clustering \citep*{takeda99}.
The current AGASA data set of 57 events above 4 $\times 10^{19}$ eV
contains four doublets and one triplet within
a separation angle of 2.5$^\circ$.
Chance probability to observe such clusters under an isotropic
distribution is about 1 $\%$ \citep*{hayashida00}.
Potential models of UHECR origin are constrained by their
ability to reproduce the observed energy spectrum and the isotropic
arrival distribution of UHECRs with the small scale clustering.

A number of studies has been devoted to the problem of the energy spectrum.
Yoshida and Teshima (1993) show the GZK cutoff in consideration
of evolution of the universe.
Blanton et al.(2001) show that the observed number of events
above $10^{20}$ eV is 2 $\sigma$ higher than the expected one,
assuming a hard $(E^{-2})$ injection spectrum and
small local overdensity (a factor of two).
Berezinsky et al.(2002) analyze three source models, GRB, AGN
and local enhancement of UHECR sources.
They conclude the AGASA events above $10^{20}$ eV, if confirmed,
must be of different origin.
Although these authors show that it is difficult to reproduce
the observed extension of the energy spectrum by the bottom-up
scenarios, they do not take into account of the effects
of the extragalactic magnetic field (EGMF) on the energy spectrum.
Furthermore, since they do not consider angular source distribution
as well as the EGMF,
the problems concerning the arrival distribution of UHECRs
can not be discussed.

Although arrival distributions from a single source are calculated
in the presence of magnetic field in some papers
\citep*{lemoine97,stanev00},
it appears to be difficult to explain the observational
isotropy by a single source.
With consideration of both the magnetic field and the source distribution,
the energy spectrum and the distribution of the arrival directions are
calculated in Sigl et al.(1999) and Lemoine et al.(1999).
However the source distribution seems to be too simple,
a two dimensional sheet \citep*{sigl99} or a pancake-like profile
\citep*{lemoine99},
which represent the Local Super Cluster (LSC).

Smialkowski et al.(2002) calculate arrival distribution
of UHECRs using the IRAS PSCz catalog of IR galaxies
as a realistic source model.
However, they restrict their attention to galaxies with far infrared
luminosity $L_{\rm {FIR}}>10^{11}L_{\odot}$
triggered by collision and merging processes,
which are possible sites of UHECR acceleration.
It is also suspected that this galaxy sample do not contain
some nearby galaxies in the LSC.
Since nearby sources within the GZK sphere are responsible for
UHECRs above $10^{20}$ eV, galaxy sample with better completeness
on nearby galaxies should be used.

Given these situations, we examined in our previous paper
\citep*{ide01}
whether the current observation can be explained by a
bottom-up scenario in which source distribution
of UHECRs is proportional to that of galaxies.
We used the Optical Redshift Survey
\citep*[ORS; ][]{santiago95} as a realistic source model.
Completeness on nearby galaxies of this sample would be
better than that of the IRAS PSCz.
By using the ORS sample, we can accurately quantify the
contribution of nearby sources to the energy spectrum above
$10^{20}$ eV.
We also took effects of the EGMF on the energy spectrum
and the arrival distribution of UHECRs into account.

However, there are problems in our previous study.
First, we did not correct absence of galaxies in the
zone of avoidance ($|b|<20^{\circ}$) when using the ORS sample,
so that we could not present accurate discussion of
statistics on the arrival directions of UHECRs.
Second, we considered propagation of UHECRs only over
40 Mpc, focusing on UHECRs with energies above
$8 \times 10^{19}$ eV.
In this case, contribution of distant sources (outside 40 Mpc)
to the cosmic ray flux below $8 \times 10^{19}$ eV
can not be taken into account.
Finally, we injected UHECRs from all the galaxies
including dwarf ones.
Luminous galaxies are generally larger than dwarf ones,
and thus they are expected to be able to accelerate
particles to higher energy.
Thus, we should investigate dependence of the results
on the limiting magnitude of galaxies.

This paper is an extended study of our earlier work.
We calculate the energy spectrum and the
arrival distribution of UHECRs using the ORS galaxy sample.
We correct the sample taking the selection effect and absence
of galaxies in the zone of avoidance ($|b|<20^{\circ}$) into account.
This allow us to accurately discuss
statistics on the arrival directions of UHECRs.
We also extend the propagation distance of UHECRs, from
40 Mpc to 1 Gpc.
As s result, we can calculate the cosmic ray flux below
$8 \times 10^{19}$ eV, properly including contribution
from distant sources.
Furthermore, we investigate dependence of the results
on the limiting magnitude of galaxies.

We compare the results of the energy spectra
with the current observations, AGASA and HiRes.
In order to explore the possibility of reproducing
the extension of energy spectrum beyond $10^{20}$ eV,
we adopt several strength of the EGMF (1,10 and 100 nG).
We also investigate dependence of the arrival distribution of
UHECRs on the limiting magnitude of galaxies and
the source number, and estimate
the most favorable ones from comparison with the AGASA data
using the harmonic analysis and the two point correlation function
as statistical methods.
We show that the three observable quantities including the GZK
cutoff of the energy spectrum can be reproduced
in the case that the number fraction $\sim 10^{-1.7}$ of the ORS
galaxies more luminous than $-20.5$ mag is selected as UHECR sources.
Implications of the results for the UHECR source candidate
are discussed in detail.

In section~\ref{model}, we show our method of calculation. 
Results are shown in section~\ref{result} and
their implications are discussed in section~\ref{discussion}.
We conclude in section~\ref{conclusion}.

\section{METHOD OF CALCULATION} \label{model}
\subsection{Method of Calculation for Propagation of UHECRs}
\label{propagation}

This subsection provides the method of Monte Carlo simulations
for propagating protons in intergalactic space.
At first, we assume that the composition of
UHECRs is proton, and inject an $E^{-2}$ spectrum
within the range of ($10^{19.5}$ - $10^{22}$)eV.
10000 protons are injected in each of 26 energy bins, that is,
10 bins per decade of energy for the cases of $l_{\rm c}$ = 40Mpc and
($B$, $l_{\rm c}$) = (1,10nG, 10Mpc), where $B$ and $l_{\rm c}$ are
strength and correlation length of the extra galactic magnetic field
(explained below).
For another cases, we inject 5000 protons in each of the bins.

Particles below $\sim 8 \times 10^{19}$ eV lose their energies
mainly by pair creations and above it by photopion production
\citep*{yoshida93} in collisions with photons of the CMB.
The pair production can be treated as a continuous loss process
considering its small inelasticity ($\sim 10^{-3}$).
We adopt the analytical fit functions given by Chododowski et al. (1992)
to calculate the energy loss rate for the pair production
on isotropic photons.
According to them, the energy loss rate of a relativistic nucleus for
the pair production on isotropic photons is given by
\begin{equation}  
- \frac{d\gamma}{dt} = \alpha r^2_0 c Z^2 \frac{m_e}{m_A}\int^{\infty}_2
d\kappa n \left( \frac{\kappa}{2\gamma}\right)
\frac{\varphi(\kappa)}{\kappa ^2}, 
\label{eqn1}
\end{equation}
where $\gamma$ is the Lorenz factor of the particle, $\kappa =  2k
\gamma$ ( $k$ is the momentum of the particle in units of $m_ec$),
$n(\kappa)$, $\alpha$, $r_0$, $Z$, and $m_A$ are the photon distribution
in the momentum space, the fine-structure constant, the classical electron
radius, the charge of the particle, and the rest mass of the particle,
respectively. 
For the energy range $E \geq 10^{19.0}$ eV, $\varphi$ can be well
represented as
\begin{equation}
\varphi \rightarrow \kappa \sum ^{3} _{i=0} d_i \ln ^i \kappa,
\label{eqn2}
\end{equation}
\begin{equation} 
d_0 \simeq -86.07,\; d_1 \simeq 50.96,\; d_2 \simeq -14.45,\;
d_3=8/3.
\label{eqn3}
\end{equation}

Since protons lose a large fraction of their energy in the
photopion production, its treatment is very important.
We use the interaction length and the energy distribution of
final protons as a function of initial proton energy
which is calculated by simulating the photopion
production with the event generator SOPHIA \citep*{sophia00}.
We checked that arrival spectrum at Earth for mono-energetic injection
of energy $E=10^{21.5}$ eV is consistent with that of
Stanev et al.(2000) (Fig. 2), which is also calculated with the
event generator SOPHIA, for the case of ($B$, $l_{\rm c}$) = (1nG,
1Mpc) which is the closest parameter set to the one of Stanev et al.(2000).

The EGMF are little known theoretically and observationally.
The upper limit for its strength and correlation length
$B_{\rm{IGM}} l_{\rm c}^{1/2} < 1 {\rm nG} (1{\rm Mpc})^{1/2}$,
as measured by Faraday rotation of radio signals from distant quasars
\citep*{kron94}, is often used \citep*{stanev00,smi02}.
We adopt in this study not only this widely accepted value, 1nG,
but also relatively strong magnetic field (10nG, 100nG),
which is better for explaining the extension of the energy spectrum
and the observed isotropic arrival distribution.
In this case, deflection angles of UHECRs become so large
that the no counterparts problem can be simply solved.

We assume that the magnetic field is represented as the Gaussian
random field with zero mean and a power-law spectrum.
Thus, $\langle B^2(k)\rangle$ can be written as
\begin{eqnarray}
\langle B^2(k)\rangle && \propto k^{n_H} \;\;\; {\rm for} \;\;\;
2 \pi / l_{\rm c} \le  k \le
2 \pi / l_{\mbox{{\scriptsize cut}}} \\
&& =0 \;\;\;\;\;\;\; \rm otherwise,
\label{eqn9}
\end{eqnarray}
where $l_{\rm c}$ is the correlation length of the EGMF
and $l_{\mbox{{\scriptsize cut}}}$ characterizes
the numerical cut-off scale.
We use $n_H=-11/3$ corresponding to the Kolmogorov spectrum.
According to our previous study \citep*{ide01},
the correlation length is chosen to be 1 Mpc, 10 Mpc, and 40 Mpc.
1 Mpc is roughly equal to mean separation of galaxies
and widely used as the typical value of the correlation length.
As we show later, the small scale clustering observed
by the AGASA is difficult to be reproduced for
$l_{\rm c} \ge 10$ Mpc (See Figure~\ref{fig13} and ~\ref{fig17}).
We do not have to consider the correlation length
larger than 40 Mpc.
Physically one expect $l_{\mbox{{\scriptsize cut}}} \ll
l_{\rm c}$, but we set $l_{\mbox{{\scriptsize cut}}} = 1/8 \times l_{\rm c}$
in order to save the CPU time.

The universe is covered with cubes of side $l_{\rm c}$.
For each of the cubes,
Fourier components of the EGMF are dialed on a cubic cell
in wave number space, whose side is $2 \pi / l_{\rm c}$,
with random phases according to the Kolmogorov spectrum,
and then Fourier transformed onto the corresponding cubic cell in real space.
We create the EGMF of 20 $\times$ 20 $\times$ 20 cubes of side $l_{\rm c}$,
and outside it,
adopt the periodic boundary condition
in order to reduce storage data for magnetic field components.
Similar methods for the turbulent magnetic fields have been
adopted \citep*{sigl99,lemoine99,isola02}.

In this study, we assume that the source distribution of UHECRs
is proportional to that of the galaxies.
We use the realistic data from the ORS \citep*{santiago95} galaxy catalog,
which is nearly full sky survey, and contains two subcatalogs,
one complete to a B magnitude of 14.5 (ORS-m) and
the other complete to a B major axis diameter of $1.'9$ (ORS-d).
Another nearly full sky catalog which is often used
in this kind of study is the IRAS PSCz Survey \citep*{blanton01,smi02}.
However, it is suspected that some galaxies in the LSC are excluded.
Since local overdensity of UHECR sources makes the GZK cutoff less sharp
or eliminate it, we use the ORS as a source model rather than the IRAS.

In any statistical acceleration mechanism, particles must
be kept confined within the acceleration site.
Luminous galaxies are generally larger than dwarf ones,
and thus they are expected to be able to accelerate particles to higher energy.
Furthermore, it is less well known that luminous galaxies in
the LSC distribute outward than faint galaxies,
contrary to general clusters of galaxies \citep*{yoshiguchi02}.
Accordingly, we inject UHECRs from galaxies with different limiting magnitudes,
and explore the source model which reproduces the current observation.
As an example, this selection of luminous galaxies
corresponds to restricting the sources of UHECRs
to host galaxies of AGNs.
It is still unknown how much an ultimate UHECR source
contribute to the observed cosmic ray flux.
We thus consider the two cases in which all galaxies are the same,
and they inject cosmic rays proportional to their absolute luminosity.
This allow us to investigate effects of the uncertainty
of the cosmic ray luminosity on the results.


\begin{figure*}
\begin{center}
\epsscale{1.2} 
\plotone{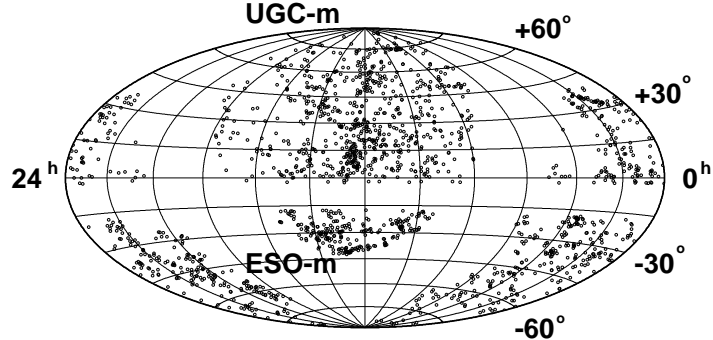} 
\caption{
Distribution of galaxies in the ESO-m and the UGC-m
samples before corrected ($v \leq 8000$ km s$^{-1}$, $M_B \leq -14.0$,
$m_B \leq 14.5$, $|b| > 20^\circ$).
\label{fig1}}
\end{center}
\end{figure*}

\vspace{0.5cm}
\centerline{{\vbox{\epsfxsize=8.0cm\epsfbox{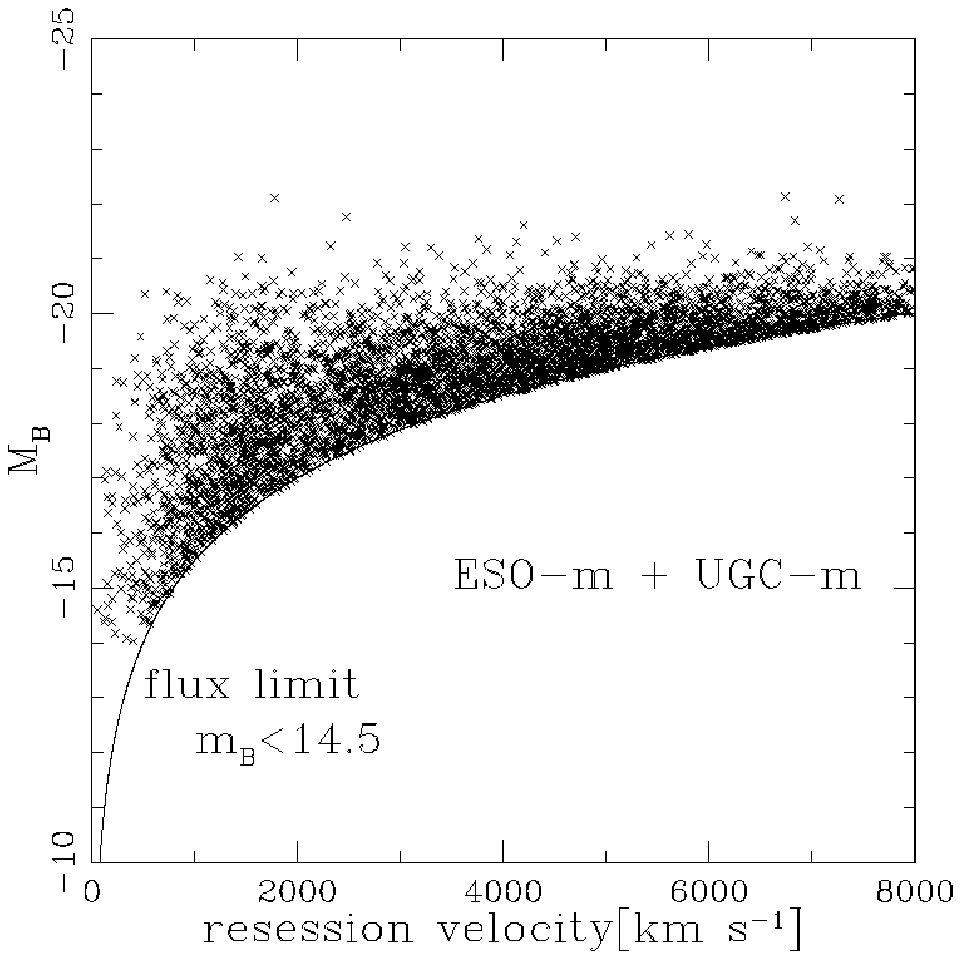}}}}
\figcaption{
Absolute magnitude plotted as a function of the recession velocity
for galaxies in our ORS sample.
\label{fig2}}
\vspace{0.5cm}

Since we consider the galaxies with different absolute magnitudes,
we use the magnitude-limited subcatalog, i.e., ORS-m, which
is also composed of two subsamples, ESO-m (2437 galaxies)
and UGC-m (3279 galaxies).
We further restrict ourselves to galaxies within 8000 km s$^{-1}$
and with $M_B - 5$log$_{10}h < -0.5 - 5$log$_{10}v$, which corresponds to
the apparent magnitude limit of $m_B=14.5$,
where $h$ is the dimensionless Hubble constant
$h=H_0/(100$ km s$^{-1}$ Mpc$^{-1})$ and $v$ is the recession velocity.
In the following, we abbreviate $M_B - 5$log$_{10}h$ to $M_B$.
We exclude very faint nearby galaxies with $M_B>-14.0$.
Galaxies in the zone of avoidance ($|b|<20^{\circ}$) are not observed. 
Our final ORS sample contains 5178 galaxies with 2346 galaxies
in the ESO-m and 2832 galaxies in the UGC-m.
Figure~\ref{fig1} shows the sky distribution of
our ORS galaxies in the equatorial coordinate. 
The strip between the ESO and UGC region $(-17.5^{\circ}
\le \delta \le -2.5^{\circ})$ is covered by the ESGC, which is a part of
the diameter-limited subcatalog, ORS-d.
The absolute magnitudes are shown in Figure~\ref{fig2}
as a function of the recession velocity.
Throughout the paper, we assume that there are no departures from
uniform Hubble expansion for simplicity, and use $h=0.75$
in calculating distance between galaxies in the sample and our Galaxy
from recession velocity.

In order to calculate the energy spectrum and the distribution of arrival
directions of UHECRs realistically, there are two key elements
of the galaxy sample to be corrected.
First, galaxies in a given magnitude-limited sample are biased tracers
of matter distribution because of the flux limit.
For each galaxy in the sample,
we compensate galaxies which is not included in the sample
using the selection function given in Santiago et al.(1996).
The positions of compensated galaxies are determined
according to Gaussian distribution whose mean is the position of the
original galaxy and rms is such that $\sigma = 1/3 \times (n \phi(r))^{-1/3}$,
where $r$ is the distance from our Galaxy and
$n$, $\phi(r)$ are number density of galaxies which is also calculated
using the selection function,
value of the selection function at distance $r$ from us respectively.
For the reason of the selection effect, the ORS is only
sampling the universe out to 8000 km s$^{-1}$ (see Figure~\ref{fig2}).
We assume that contributions from sources outside it is completely
isotropic, and calculate its amount from the number of galaxies inside it.
Second, our ORS sample does not include galaxies
in the zone of avoidance ($|b|<20^{\circ}$) and the
ESGC region $(-17.5^{\circ} \le \delta \le -2.5^{\circ})$.
In the same way, we assume that the source distribution in this region is
homogeneous, and calculate its number density from the number of
galaxies in the observed region, which is corrected for the selection effect.
In Figure~\ref{fig3}, we show the sky distribution of the ORS galaxies
corrected in the manner explained above.

\centerline{{\vbox{\epsscale{1.0}\plotone{fig3.eps}}}}
\figcaption{
Distribution of galaxies in our ORS sample within 8000 km s$^{-1}$,
which are corrected for the selection effect in the manner explained
in the text.
\label{fig3}}
\vspace{0.5cm}

Finally, we now explain how the angular probability
distribution of UHECRs are calculated.
At first, protons with a flat energy spectrum are injected isotropically
at a given point in each of the 26 energy bins, and then propagated
in the EGMF over $1$ Gpc for $15$ Gyr.
Weighted with a factor corresponding to a $E^{-2}$ power law spectrum,
this provides distribution of energy, deflection angle, and time delay
of UHECRs as a function of the distance from the initial point.
With this distribution, we can calculate the flux and the arrival
directions of UHECRs injected at a single UHECR source.
Then, summing contributions from all the sources,
we obtain the angular probability distributions of UHECRs.
In this paper, we use the distribution of energies and deflection angles
integrated over the time delay, assuming that the cosmic ray flux
at the earth is stationary.
In Figure~\ref{fig4}, we show number of protons with energies of
$(10^{19.6}-10^{20.3})$ eV integrated over deflection angle,
time delay as a function of propagation distance.
It is noted that this number is proportional to the cosmic ray
flux in this energy range from sources in concentric shells with an
infinitely small width, assuming that a source distribution is homogeneous.

\vspace{0.5cm}
\centerline{{\vbox{\epsfxsize=7.5cm\epsfbox{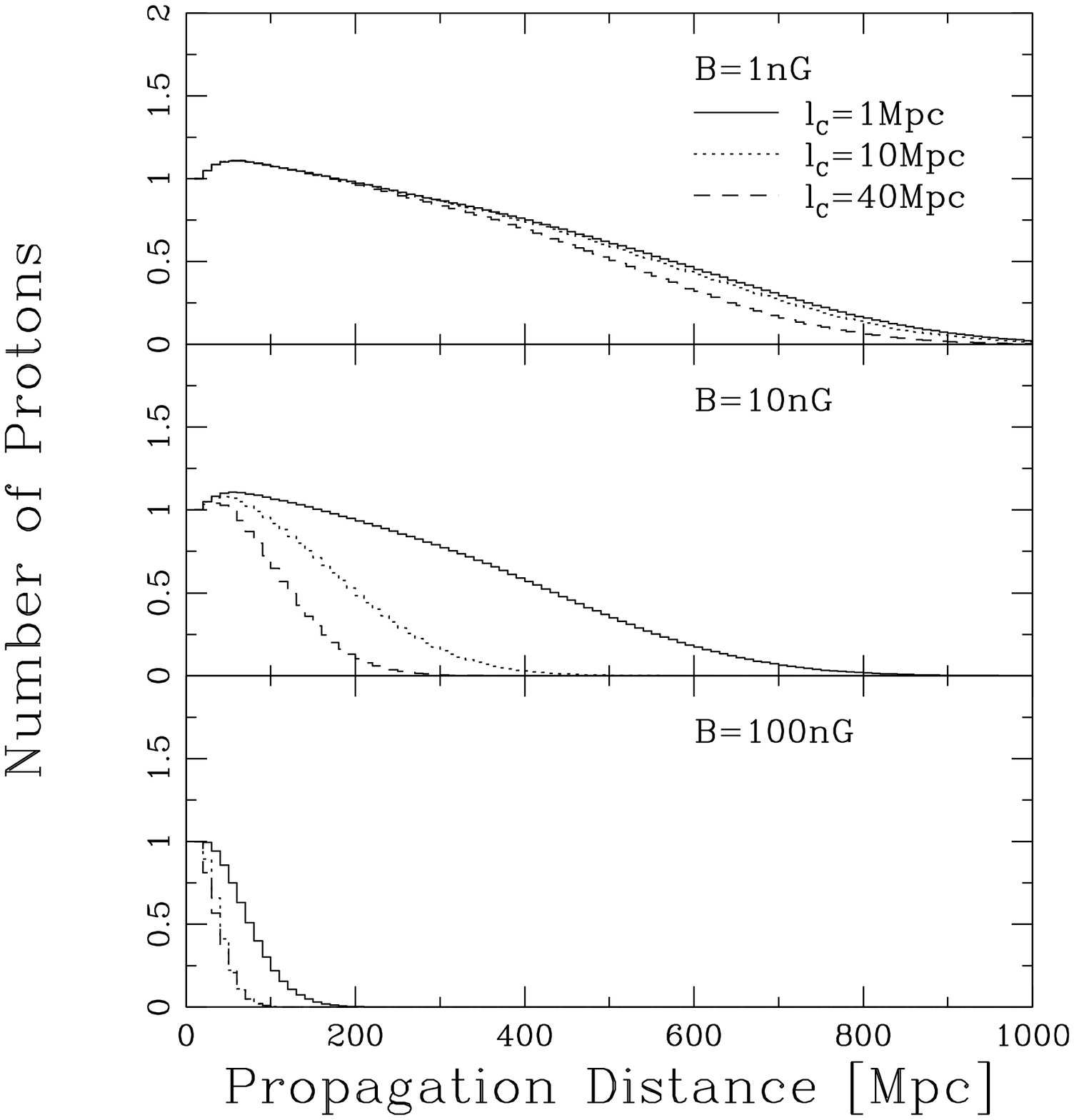}}}}
\figcaption{
Number of protons with arrival energies of $(10^{19.6}-10^{20.3})$ eV
integrated over deflection angle and time delay after propagation from an
initial point as a function of the propagation distance.
The number is normalized to be 1 at the injection.
It is noted that this number is proportional to the cosmic ray
flux in this energy range from sources in concentric shells with an
infinitely small width, assuming that a source distribution is homogeneous.
\label{fig4}}
\vspace{0.5cm}

\subsection{Statistical Methods}\label{statics}

In this subsection, we explain the two statistical quantities,
the harmonics analysis for large scale anisotropy \citep*{hayashida99}
and the two point correlation function for small scale anisotropy.
The harmonic analysis to the right ascension distribution of events
is the conventional method to search for global anisotropy
of cosmic ray arrival distribution.
For a ground-based detector like the AGASA, the almost uniform
observation in right ascension is expected.
The $m$-th harmonic amplitude $r$ is determined by fitting the distribution
to a sine wave with period $2 \pi /m$.
For a sample of $n$ measurements of phase,
$\phi_1$, $\phi_2$, $\cdot \cdot \cdot$, $\phi_n$
(0 $\le \phi_i \le 2 \pi$), it is expressed as
\begin{equation}
r = (a^2 + b^2)^{1/2}
\label{eqn121}
\end{equation}
where, $a = \frac{2}{n} \Sigma_{i = 1}^{n} \cos m \phi_i  $,
$b = \frac{2}{n} \Sigma_{i = 1}^{n} \sin m \phi_i  $.
We calculate the first and second amplitude from a set of events
generated according to predicted probability density distribution
of arrival directions of UHECRs.

If events with
total number $n$ are uniformly distributed in right ascension, the chance
probability of observing the amplitude $\ge r$ is given by,
\begin{equation}
P = \exp (-k),
\label{eqn13}
\end{equation}
where
\begin{equation}
k = n r^2/4.
\label{eqn14}
\end{equation}
The AGASA 57 events is consistent with isotropic source distribution
within 90 $\%$ confidence level as shown later.
Accordingly, we adopt the amplitude such that $P = 0.1$
as observational constraint, rather than the most probable
amplitude calculated from the observed 57 events.

The two point correlation function $N(\theta)$ contains
information on the small scale anisotropy.
We start from a set of generated events or the actual AGASA data.
For each event, we divide the sphere into concentric bins of
angular size $\Delta \theta$, and count the number of events falling
into each bin.
We then divide it by the solid angle of the corresponding bin,
that is,
\begin{eqnarray}
N ( \theta ) = \frac{1}{2 \pi | \cos \theta  - \cos (\theta + \Delta \theta)
|} \sum_{ \theta
\le  \phi \le \theta + \Delta \theta }  1 \;\;\; [ \rm  sr ^{-1} ],
\label{eqn100}
\end{eqnarray}
where $\phi$ denotes the separation angle of the two events.
$\Delta \theta$ is taken to be $1^{\circ}$ in this analysis.
The AGASA data shows strong correlation at small angle $(\sim 2^{\circ})$
with 5 $\sigma$ significance of deviation from an isotropic distribution
as shown later.

In order to quantify the statistical significance of deviations
between models and data we introduce $\chi_{ \theta_{\rm {max}} }$,
which is defined by,
\begin{eqnarray}
\chi_{ \theta_{\rm {max}} } = \frac{1}{\theta_{\rm {max}}} \sqrt{
\sum^{ \theta_{\rm {max}} }_{ \theta_{\rm i} =0 }
\frac{ \left( N ( \theta_{\rm i} ) - N_{\rm {obs}}
( \theta_{\rm i} )\right) ^2 }
{ \sigma_{\rm i} ^2 }  },
\label{eqnchi}
\end{eqnarray}
where $N_{\rm {obs}} ( \theta_{\rm i} )$ is the value of
the two point correlation
function obtained by the AGASA observation at angle $\theta_{\rm i}$,
and $\sigma_{\rm i}$ is statistical error of the $N ( \theta_{\rm i} )$
due to the finite number of simulated events,
which is set to be the observed one.
In this study, we take $\theta_{\rm {max}}$ to be 10 deg for
properly quantifying deviations of models from the sudden
increase of correlation at the small angle scale observed by the AGASA.

We compare the model predictions with the existing AGASA data
throughout the paper.
Therefore, we set the number of simulated events to be that of the AGASA data
above $10^{19.6}$ eV (57), and restrict the arrival directions of UHECRs
in the range of $-10^{\circ} \le \delta \le 80^{\circ}$
when calculating these statistical quantities.
There are also the present working or development of large-aperture
new detectors, such as the HiRes
\citep*{wilkinson99} and South and North Auger \citep*{capelle98}.
If they report the arrival directions of observed events,
we can make comparisons between the model predictions and
their data using the same statistical methods.

\section{RESULTS} \label{result}
\subsection{Angular Images of UHECRs}\label{image}
\indent

In this subsection, we present the results of the numerical simulations
for the angular images of UHECRs
in a variety of conditions concerning strength and correlation length
of the EGMF and the limiting magnitudes of galaxies.
We show the angular images for
two energy range $(10^{19.6}-10^{20})$ eV and above $10^{20}$ eV,
about at which the GZK cutoff of the energy spectrum is predicted.
For comparison, we also show the AGASA events
with the angular images.
It is noted that the current AGASA data set contains
49 events with energies of $(10^{19.6}-10^{20})$ eV
and 8 events above $10^{20}$ eV.

\begin{figure*}
\begin{center}
\epsscale{1.4} 
\plotone{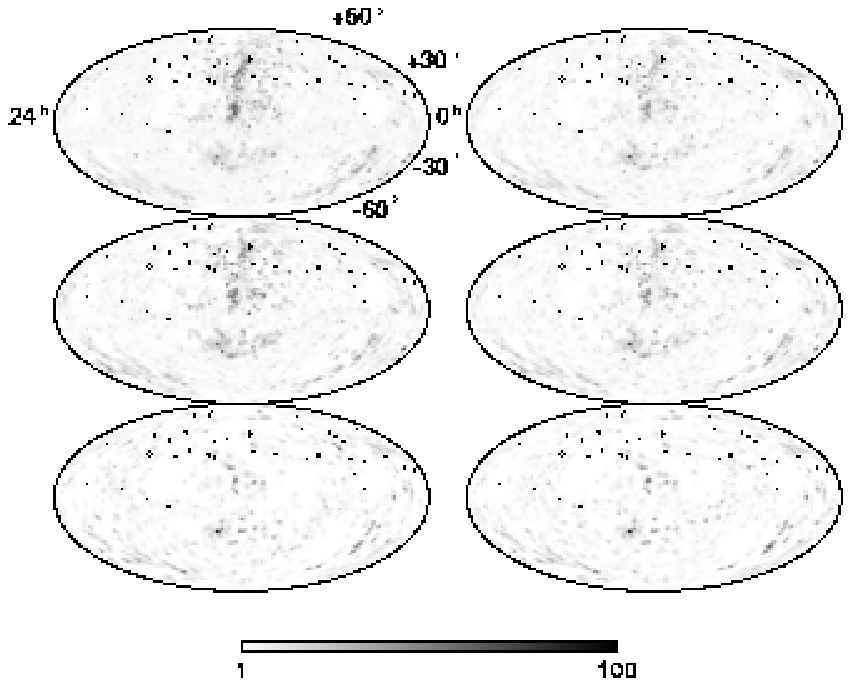} 
\caption{
Angular probability distribution of UHECRs with
energies of $(10^{19.6}-10^{20.0})$ eV
originating the galaxies more luminous than $M_{\rm{lim}}=-14.0$ (upper),
$-18.0$ (middle), $-20.0$ (bottom) for number (left) and luminosity (right)
weighted sources in the case of $(B,l_{\rm c})=(1,1)$.
Resolution of the image is set to be $1^{\circ} \times 1^{\circ}$.
The AGASA 49 events in the corresponding energy range are also shown
as circles of radius proportional to their energies.
\label{fig5}}
\end{center}
\end{figure*}

The calculated angular images of UHECRs
originating the galaxies more luminous than $M_{\rm{lim}}=-14.0$ (upper),
$-18.0$ (middle), $-20.0$ (bottom) for number (left) and
luminosity (right) weighted sources
in the case of $(B,l_{\rm c})=(1,1)$ are shown in Figure~\ref{fig5}
for energy range $(10^{19.6}-10^{20})$ eV.
The AGASA 49 events in this energy range are also shown
as circles of radius proportional to their energies.
Spatial structure of the LSC is reflected by prominent
high intensity region running north and south at $\alpha \sim 180^{\circ}$
for $M_{\rm{lim}}=-14.0$ and $-18.0$.
Since the luminous galaxies which distribute outward than the faint ones
make a substantial contribution in the case of luminosity weighted sources,
the region of high intensity becomes to be less obvious
in the case of $M_{\rm{lim}}=-14.0$.
In other cases, there is no significant difference simply due to
small variance of absolute magnitudes of the individual galaxy.
We can see that the angular image becomes to be isotropic and
show good correlation with AGASA events
as restricting sources to more luminous galaxies.
Such correlation of the AGASA events with angular image of UHECRs
is also pointed out in the case of galaxies with huge infrared luminosity
\citep*{smi02}.

Even for $M_{\rm{lim}}=-20.0$, there are high intensity regions
around $(\alpha, \delta) \sim (180^{\circ}, 10^{\circ})$
which do not correlate with the AGASA events.
However, if we select sources, which contribute to the currently
observed cosmic ray flux, from our ORS sample,
these high intensity regions can be eliminated.
An example in the case of $M_{\rm{lim}}=-20.5$
is shown in Figure~\ref{fig21} as we explain later.

\begin{figure*}
\begin{center}
\epsscale{1.5}
\plotone{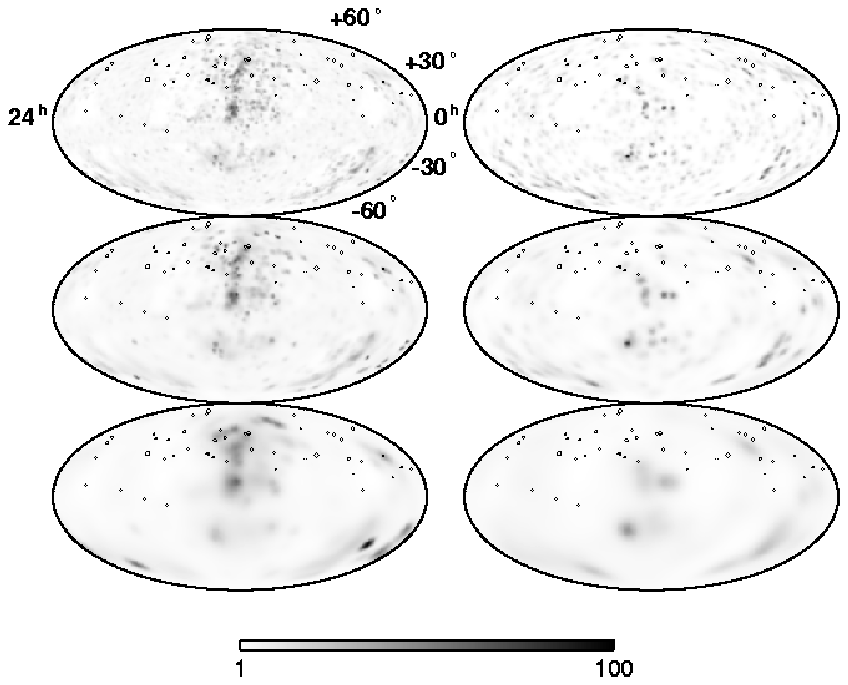} 
\caption{
Same as Figure 5, but for $(B,l_{\rm c})=(1,1)$ (upper), $(1,10)$ (middle),
$(10,1)$ (bottom) and $M_{\rm{lim}}=-14.0$ (left), $-20.0$ (right)
in the case of number weighted sources.
\label{fig6}}
\end{center}
\end{figure*}

Figure~\ref{fig6} shows the predicted angular image of UHECRs
for $(B,l_{\rm c})=$(1,1) (upper), (1,10) (middle), (10,1) (bottom)
and $M_{\rm{lim}}=-14.0$ (left), $-20.0$ (right)
with energies of $(10^{19.6}-10^{20})$ eV.
The angular image is distorted for $(B,l_{\rm c})=$(1,10), (10,1),
and dependence on the correlation length is relatively weak
because the deflection angle of UHECRs is proportional
to $B \cdot l_{\rm c}^{1/2}$.
As we will show later, this distortion of the angular image
is crucial for the small scale clustering.

\begin{figure*}
\begin{center}
\epsscale{1.5}
\plotone{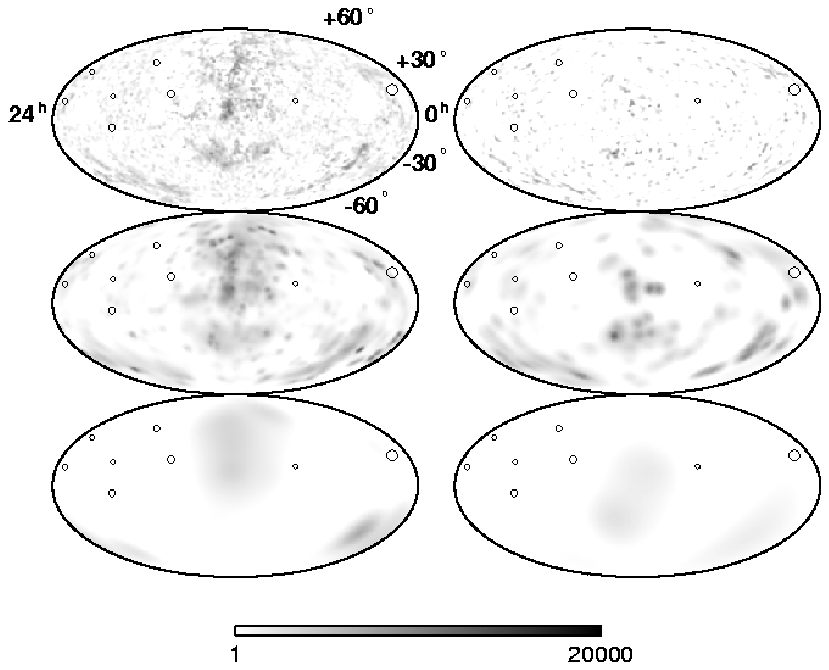} 
\caption{
Same as Figure 5, but with energy of $(10^{20}-10^{20.3})$ eV
for $(B,l_{\rm c})=(1,1)$ (upper), $(10,1)$ (middle), $(100,1)$ (bottom)
and $M_{\rm{lim}}=-14.0$ (left), $-20.0$ (right).
The AGASA 8 events above $10^{20}$ eV are also shown
as circles of radius proportional to their energies.
\label{fig7}}
\end{center}
\end{figure*}

In Figure~\ref{fig7}, we show the angular image
for $(B,l_{\rm c})=(1,1)$ (upper), $(10,1)$ (middle), $(100,1)$ (bottom)
and $M_{\rm{lim}}=-14.0$ (left), $-20.0$ (right)
with the energy range $(10^{20}-10^{20.3})$ eV.
(Highest energy of the events observed by the AGASA is $\sim 10^{20.3}$ eV.)
Because protons with energies above $10^{20}$ eV must originate
within about 50 Mpc from us due to photopion production,
only nearby galaxies are able to make a substantial contribution
to the angular image so that the spatial structure of the LSC
is clearly visible.
There is no correlation for $M_{\rm{lim}}=-14.0$
between the AGASA events and the angular image of UHECRs.
For $M_{\rm{lim}}=-20.0$, arrival distribution is relatively
isotropic, but still has no correlation with the AGASA events.
As we conclude later, the AGASA events above $10^{20}$ eV
might be of different origin, if they are confirmed.

\subsection{Energy Spectra of UHECRs}\label{spectrum}
\indent

We present in this subsection the results of the numerical
simulations for the energy spectra of UHECRs.
Figure~\ref{fig8} shows the energy spectrum in the case of
$B=1$ (upper), $10$ (middle), $100$ (bottom) nG and
$M_{\rm{lim}}=-20.0$ for number weighted sources.
They are normalized arbitrary, but with the same factors for
all the values of the correlation length.
The substantial deflection in the EGMF results in a significant
increase of the path length, then leading to a time delay exceeding
the age of the universe or a decrease of energy below $E=10^{19.6}$ eV
due to pair production for protons injected at sufficiently large distance,
as easily seen in Figure~\ref{fig4}.
Thus, the stronger EGMF make the GZK cutoff around $10^{20}$ eV
less sharp or eliminate it in Figure~\ref{fig8}.

\vspace{0.5cm}
\centerline{{\vbox{\epsfxsize=7.5cm\epsfbox{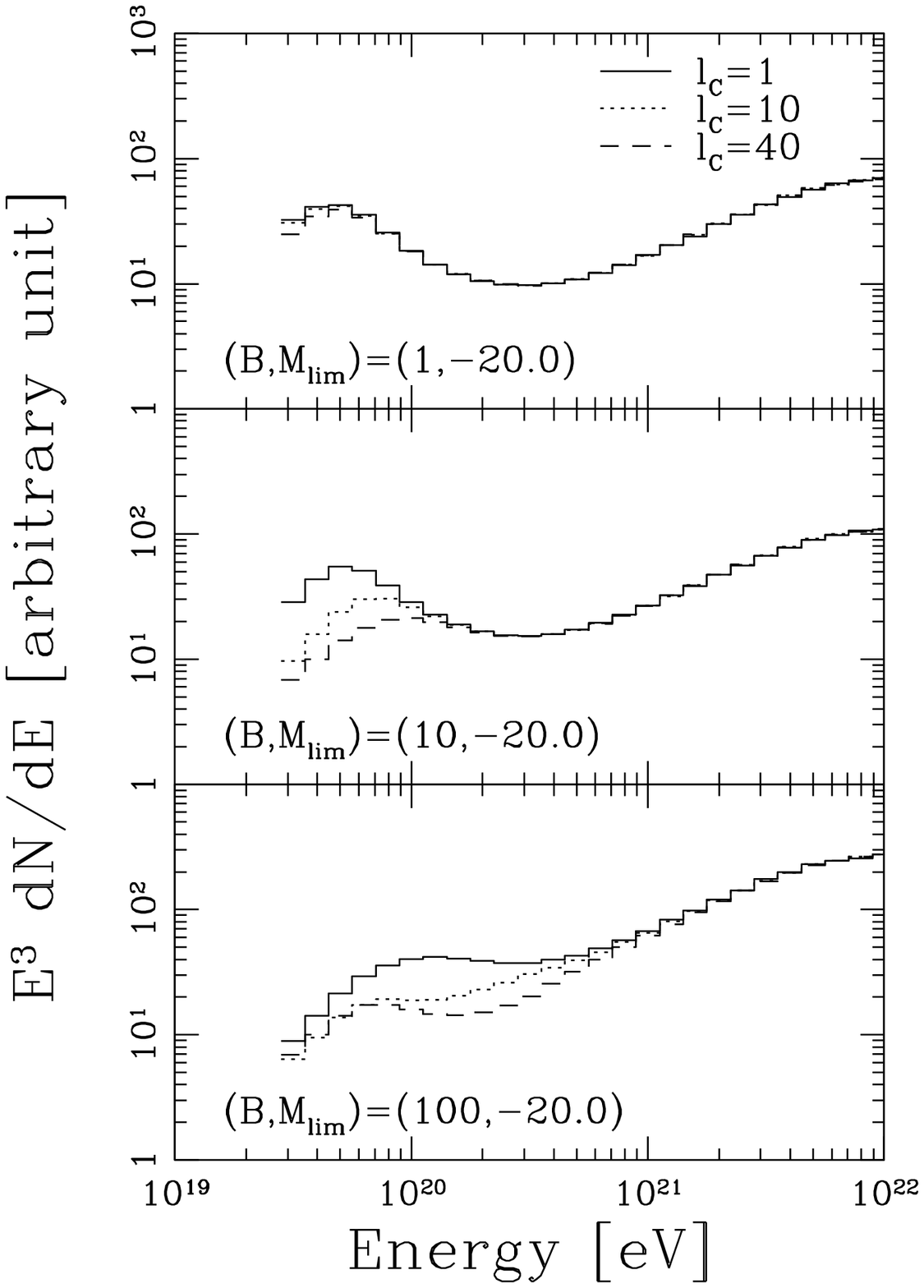}}}}
\figcaption{
The energy spectra in the case of
$B=1$ (upper), $10$ (middle), $100$ (bottom) nG and
$M_{\rm{lim}}=-20$ for number weighted sources.
They are normalized arbitrarily, but with the same factors
for all the values of the correlation length.
\label{fig8}}
\vspace{0.5cm}

When the correlation length of the EGMF is small,
protons can not be deflected effectively and thus
its path length and equivalently energy loss is smaller than
that for longer correlation length.
This can also be seen in Figure~\ref{fig4} and Figure~\ref{fig8}.
Especially, effects of changing the correlation length
are large at $E < 10^{20}$ eV for $B=10$ nG and 
at $E \sim 10^{20}$ eV for $B=100$ nG,
where gyroradii of UHECRs are about $1$ Mpc and thus
deflection is strong enough for the diffusion approximation
to become applicable in the case of $l_{\rm c}=10$ and $40$ Mpc.

In order to determine normalization of the energy spectrum
and quantify statistical significance of mean deviation between
the fitted energy spectrum and the observed one,
we introduce $\chi_{ES}$ in the similar way as
the two point correlation function explained in the previous section.
This is defined by
\begin{eqnarray}
\chi_{ \rm {ES} } = \frac{1}{9} \sqrt{
\sum_{ E_{\rm i} = 10^{19.5} }^{ 10^{20.3} }
\frac{ \left( dN/dE ( E_{\rm i} ) - dN_{\rm {obs}}/dE ( E_{\rm i} )\right)
^2 } { \sigma_{\rm i} ^2 }  },
\label{eqnchies}
\end{eqnarray}
where $dN/dE_{\rm {obs}} ( E_{\rm i} )$ is the energy spectrum observed
by the AGASA at $E=E_{\rm i}$, and $\sigma_{\rm i}$ is $1 \sigma$ error
at this energy bin.

The origin of cosmic rays with energies below $10^{19.5}$ eV
is also one of the major open questions in astro-particle physics.
Berezinsky et al.(2002) showed that predicted UHECRs flux from GRBs
fall short of the observed flux below $10^{19.5}$ eV assuming an
injection spectrum $E^{-2}$, which is advocated in Vietri (1995)
and Waxman (1995), and also used in this study.
However, there are another possible extra-galactic and galactic
UHECR production sites, such as AGNs and young neutron stars
\citep*{blasi00}.
Maximum energy of cosmic ray achieved by them may be lower than $10^{19.5}$ eV.
In this case,
these galactic and/or extra-galactic components may substantially
contribute to the cosmic ray flux below $10^{19.5}$ eV.
Throughout the paper, we assume that these components
bridge the gap between the observed flux
and the predicted one of Berezinsky et al.(2002),
and restrict ourselves to the energy spectrum only above $10^{19.5}$ eV.

\vspace{0.5cm}
\centerline{{\vbox{\epsfxsize=7.5cm\epsfbox{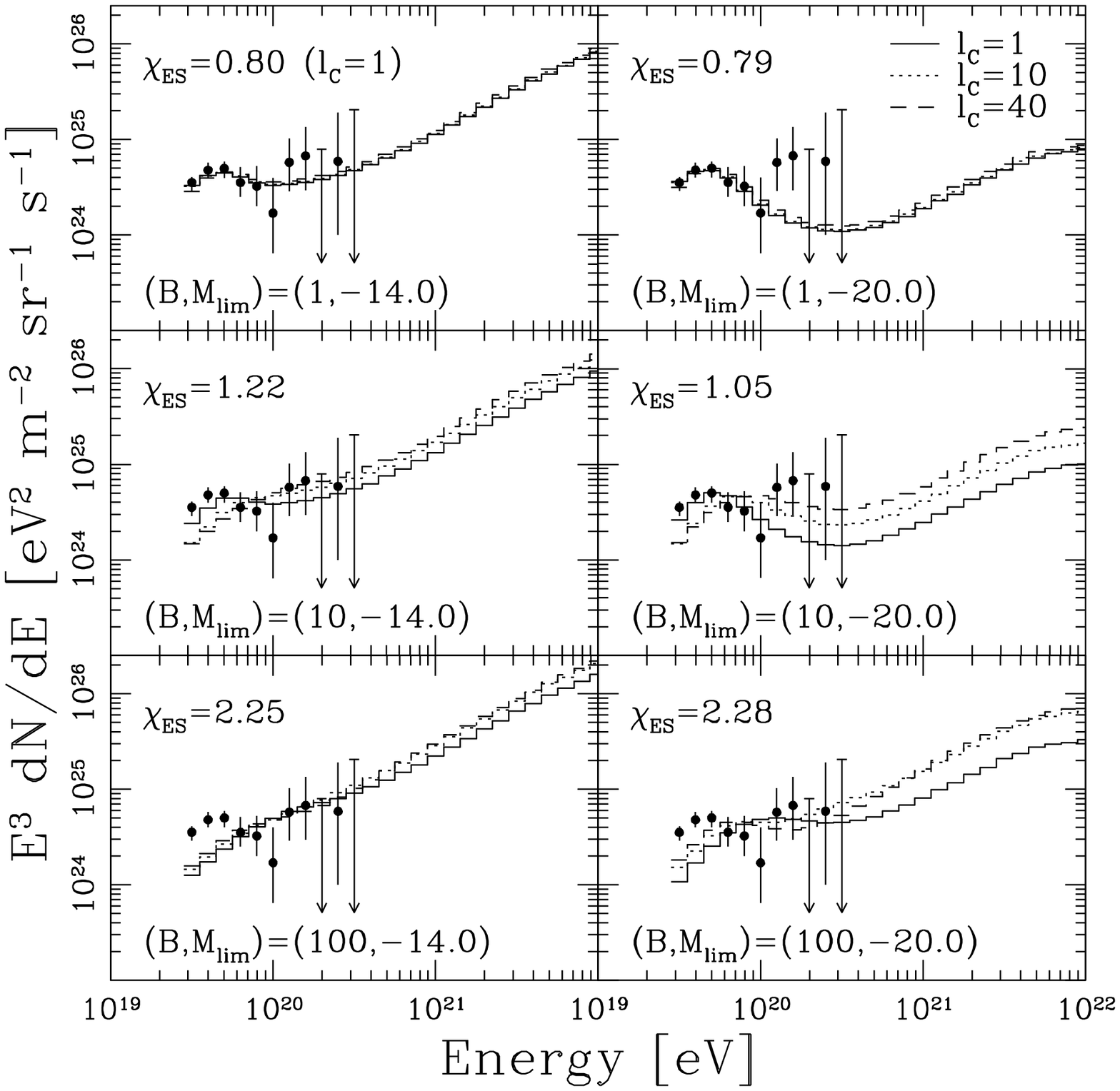}}}}
\figcaption{
The energy spectra in the case of
$B=1$ (upper), $10$ (middle), $100$ (bottom) nG and
$M_{\rm{lim}}=-14$ (left), $-20.0$ (right) for number weighted sources.
They are normalized so as to minimize
$\chi_{\rm {ES}}$, which is shown in this figure
for $l_{\rm c}=1$ Mpc.
The energy spectrum observed by the AGASA is also shown
\citep*{hayashida00}.
\label{fig9}}
\vspace{0.5cm}

Figure~\ref{fig9} shows the energy spectrum normalized so as
to minimize $\chi_{ES}$
in the case of $B=1$ (upper), $10$ (middle), $100$ (bottom) nG and
$M_{\rm{lim}}=-14.0$ (left), $-20.0$ (right) for number weighted sources.
We also show the significance values $\chi_{ES}$ in this figure.

\vspace{0.5cm}
\centerline{{\vbox{\epsfxsize=7.5cm\epsfbox{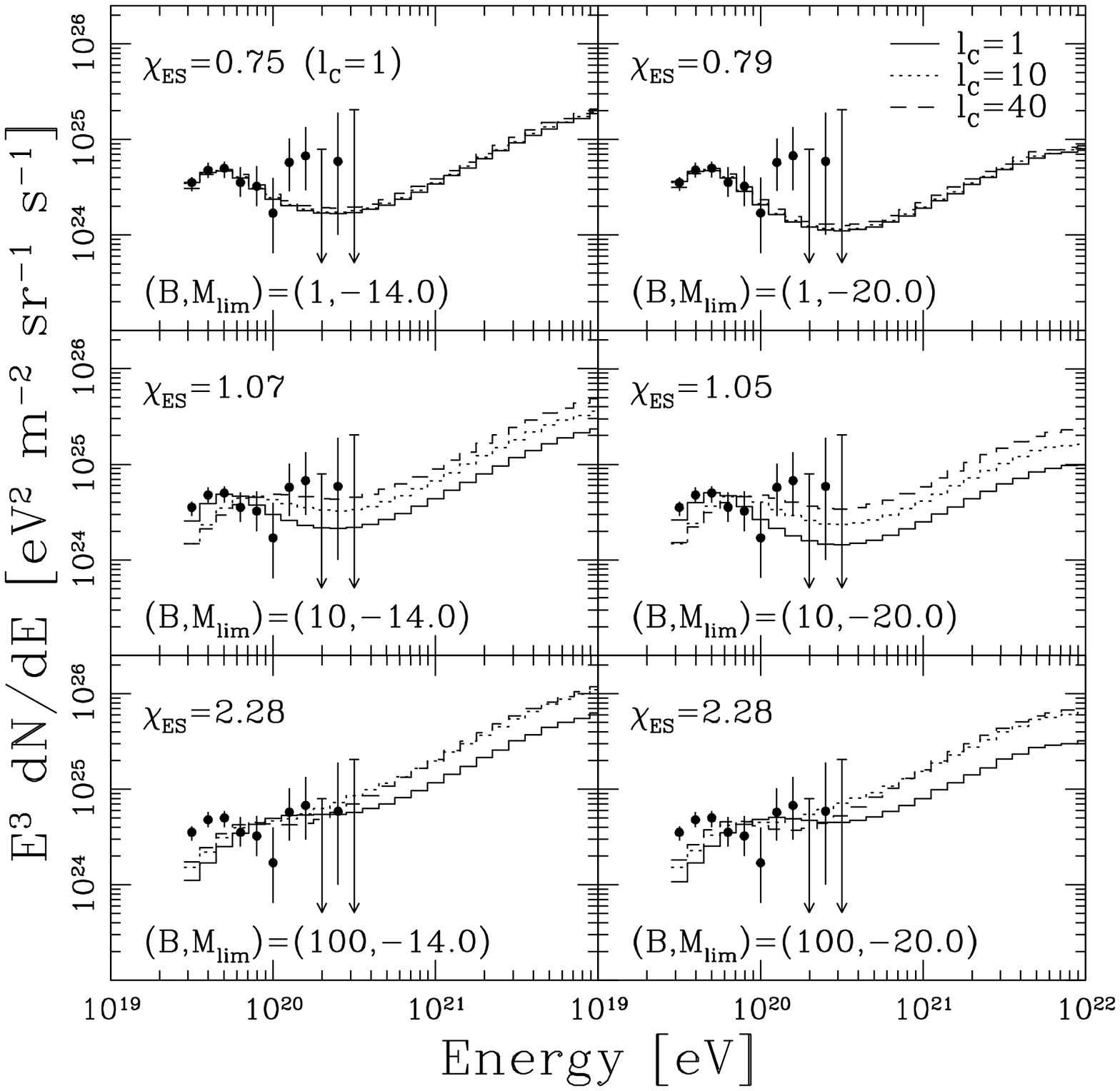}}}}
\figcaption{
Same as Figure~\ref{fig9}, but for luminosity weighted sources.
\label{fig10}}
\vspace{0.5cm}

The predicted energy spectrum for $(B,M_{\rm {lim}})=(1,-14.0)$
provide good fit to the AGASA data, including the extension above
$10^{20}$ eV.
However, arrival distribution is significantly anisotropic
and inconsistent with the AGASA (see the next subsection).
In the case of $(B,M_{\rm {lim}})=(1,-20.0)$, the energy spectrum can be well
fitted with the AGASA data below $E=10^{20}$ eV,
but the GZK cutoff is predicted above $E=10^{20}$ eV due to the energy loss
by photopion production in contrast to the AGASA data.
This is because luminous galaxies near the earth are less than dwarf galaxies
(see Figure~\ref{fig2}).
Although the energy spectrum extends beyond $10^{20}$ eV
for the stronger EGMF, it does not provide good fit
with the AGASA data at $E<10^{20}$ eV, where the current data
have adequate statistics.
Consequently, $\chi_{ES}$ for the stronger EGMF is larger than that
for $B=1$ nG as easily seen in Figure~\ref{fig9}.

In the same way as Figure~\ref{fig9}, we show the energy spectra
for luminosity weighted sources in Figure~\ref{fig10}.
Since contributions from the dwarf galaxies are suppressed in this case,
the cutoff of the energy spectra for $M_{\rm{lim}}=-14.0$
is predicted.
For $M_{\rm{lim}}=-20.0$, the results are almost the same as number
weighted ones.

\subsection{Statistics on the Arrival Directions of UHECRs}\label{statisresult}

In this subsection, we show the results of statistics
on the arrival directions of UHECRs.
Figure~\ref{fig11} shows the predicted first harmonics for
number weighted sources as a function of limiting magnitude.
The error bars represent the statistical error due to the
finite number of simulated events, whose number is set to be
that observed by the AGASA in the energy range of $(10^{19.6}-10^{20.3})$ eV.
Arrival directions of UHECRs are restricted in the range
$-10^{\circ} \le \delta \le 80^{\circ}$ in order to compare our results
with the AGASA data.
The solid lines indicate the harmonic amplitude obtained by the AGASA data.
The shaded region is expected from the statistical fluctuation
of isotropic source distribution with the chance probability
larger than $10 \%$.

\vspace{0.5cm}
\centerline{{\vbox{\epsfxsize=7.5cm\epsfbox{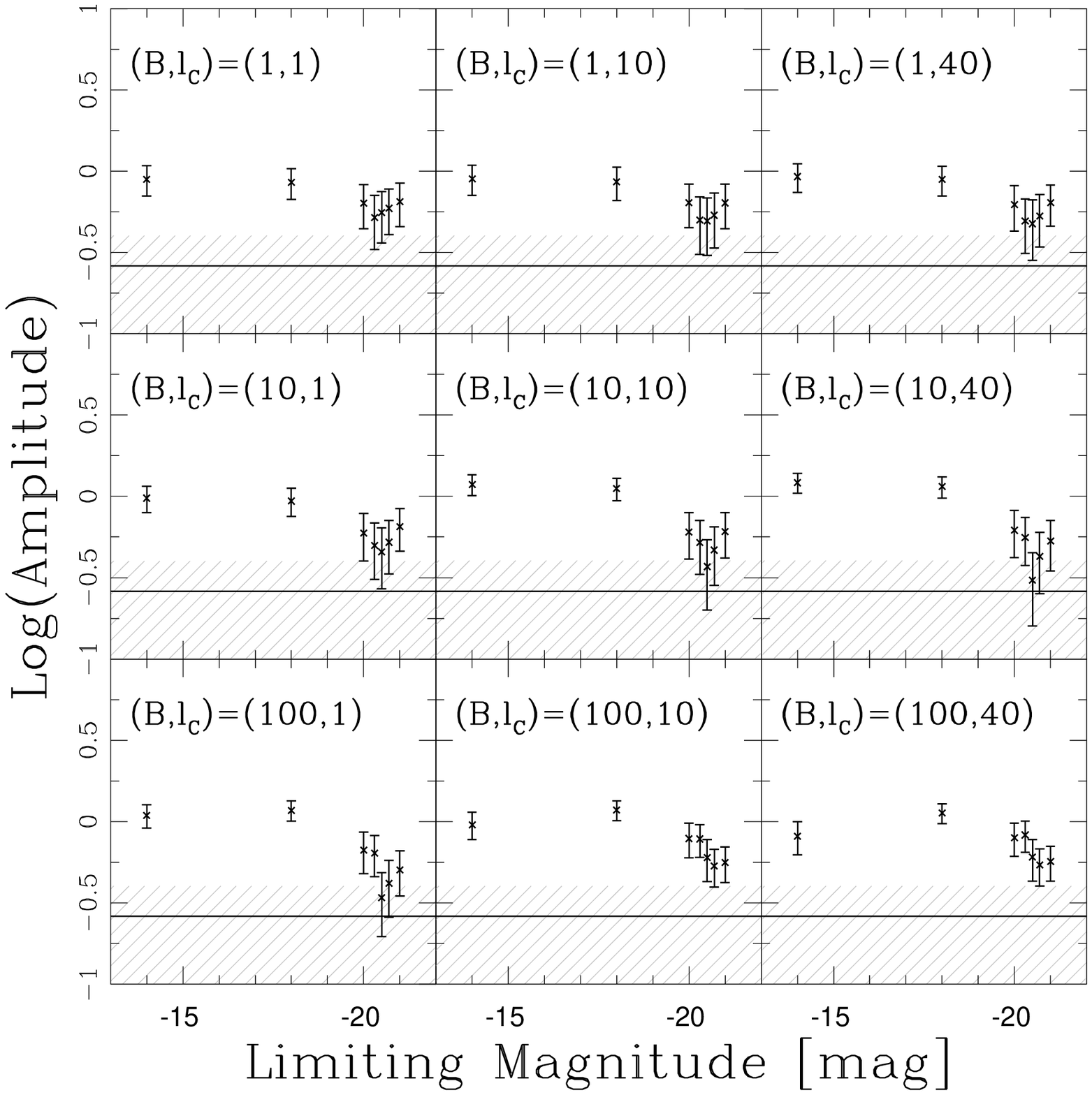}}}}
\figcaption{
Amplitude of the first harmonics as a function of the limiting
magnitude of galaxies in the case of number weighted sources.
The number of simulated events is set to be 57 in the energy range of
$(10^{19.6}-10^{20.3})$.
We dial sets of events 1000 times from the calculated
probability density distribution in order to obtain averages and variances
of the amplitude.
The solid lines represent the AGASA data in the corresponding energy range.
The shaded region is expected from the statistical fluctuation
of isotropic source distribution with the chance probability
larger than $10 \%$.
\label{fig11}}
\vspace{0.5cm}

Clearly visible in Figure~\ref{fig11} is that the anisotropy
is least for $M_{\rm{lim}} \sim -20.5$, which is caused by broadened
distribution of luminous galaxies in the LSC.
The amplitude at $M_{\rm {lim}}=-20.5$ is
consistent with isotropic source distribution within 1 $\sigma$ level
for all the $(B,l_{\rm c})$.
Compared with the AGASA data, a significant anisotropy can be seen
for $M_{\rm{lim}} \ge -18.0$, although the energy spectra in these cases
extend above $10^{20}$ eV as shown in the previous subsection.
Increase of the amplitude around $M_{\rm{lim}} \sim -21.0$ is mainly due to
two giant galaxies existing in the almost same direction
$(\alpha \sim 185^{\circ}, \delta \sim 14.7^{\circ})$
in the vicinity of our Galaxy (at $\sim$ 30 Mpc).

Comparing the results of different strength of the EGMF,
we find that the anisotropy in the cases of 10, 100 nG is larger
than that of 1 nG below $M_{\rm{lim}} \sim -18.0$ contrary to expectation.
We note again that stronger EGMF results in shorter cosmic ray horizon
because of time delay and energy loss due to
the substantial deflection in the EGMF (See Figure~\ref{fig4}).
That is, spatial structure of nearby galaxies is strongly reflected
in the large scale anisotropy for the strong EGMF.
Above $M_{\rm{lim}} \sim -18.0$, since luminous galaxies distribute outward,
the anisotropy in the cases of the strong EGMF is smaller as expected.
Also, increase of correlation length results in more effective deflections,
and then leads to smaller anisotropy in the case of 1, 10 nG.
However, this dependence is opposite for 100 nG, which is caused by a reason
somewhat similar to that of slightly large anisotropy for the strong EGMF
below $M_{\rm{lim}} \sim -18.0$ explained above.
Gyroradius of proton with energy $10^{20.0}$ eV in the EGMF
of 100 nG is 1 Mpc, which is smaller than $l_{\rm c}=10, 40$ Mpc.
In this case, UHECR deflection is strong enough for the diffusion
approximation to become applicable.
Thus, the cosmic ray horizon is extremely shorter
than that for $l_{\rm c}=1$ Mpc,
so that the anisotropy arise from nearby galaxies in the LSC.

Figure~\ref{fig12} shows the predicted second harmonics for
number weighted sources in the same manner as the first harmonics.
Dependence on the limiting magnitude is weaker than
that of the first harmonics.
There is no parameter set $(B,l_{\rm c},M_{\rm{lim}})$ which is
consistent with isotropic source distribution.
We also calculated the first and second harmonics for luminosity weighted
sources, and found that the anisotropy is slightly less than
that of number weighted sources for $M_{\rm{lim}} \geq -18.0$
because of small contributions from the dwarf galaxies,
which concentrate in the LSC.
For $M_{\rm{lim}} < -18.0$, the results are almost the same
as number weighted sources.

\vspace{0.5cm}
\centerline{{\vbox{\epsfxsize=7.5cm\epsfbox{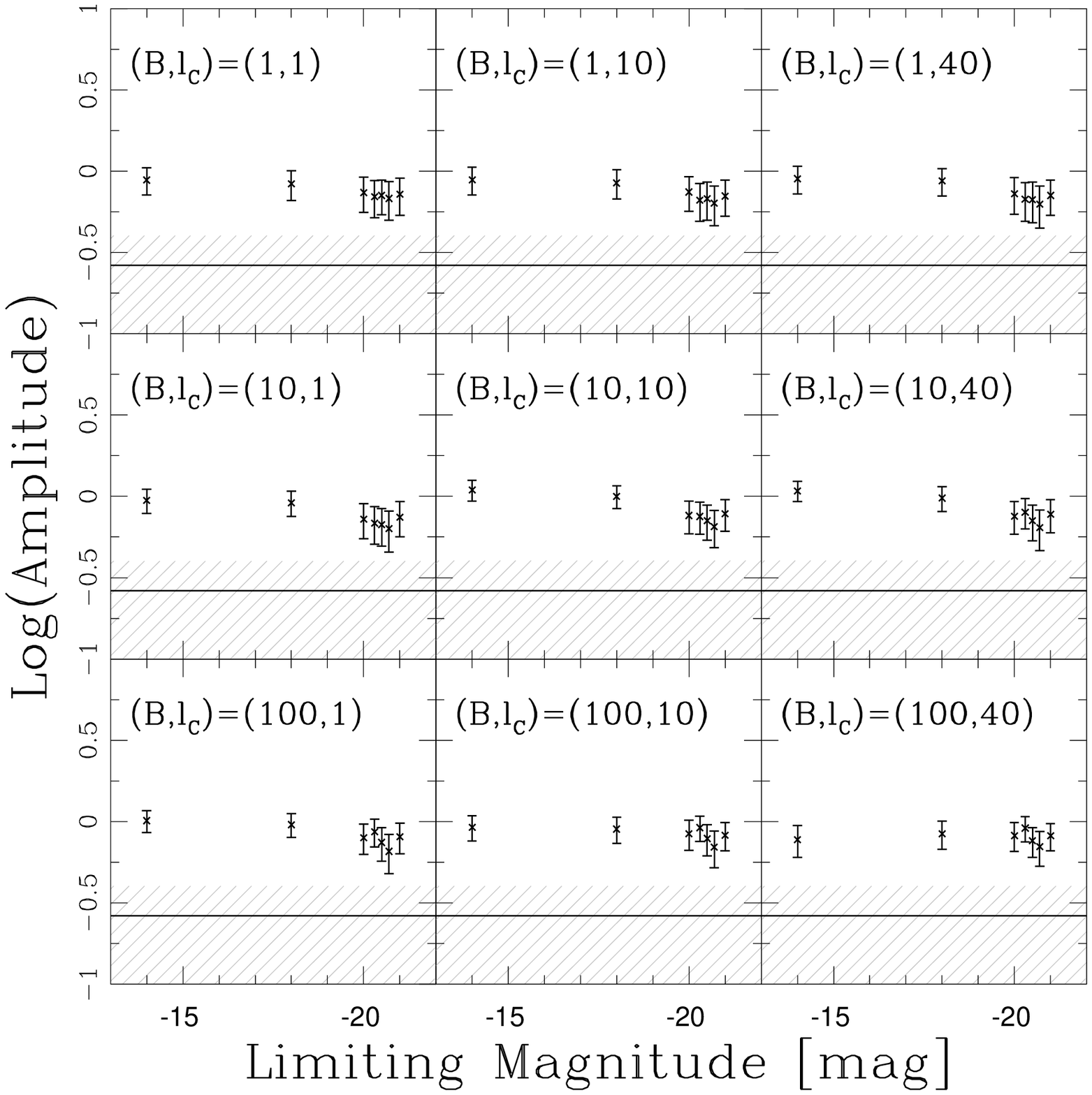}}}}
\figcaption{
Same as Figure 11, but amplitude of the second harmonics.
\label{fig12}}
\vspace{0.5cm}

The two point correlation functions are shown
for number weighted sources in the case of $M_{\rm{lim}} = -14.0$ (left panels)
and $M_{\rm{lim}} = -20.5$ (right panels) in Figure~\ref{fig13}.
The histograms represent the AGASA data, which show the statistically
significant correlation at the smallest angle scale $(\sim 1^{\circ})$.
$\chi_{ 10 }$ defined in the previous section is also shown.

\vspace{0.5cm}
\centerline{{\vbox{\epsfxsize=7.5cm\epsfbox{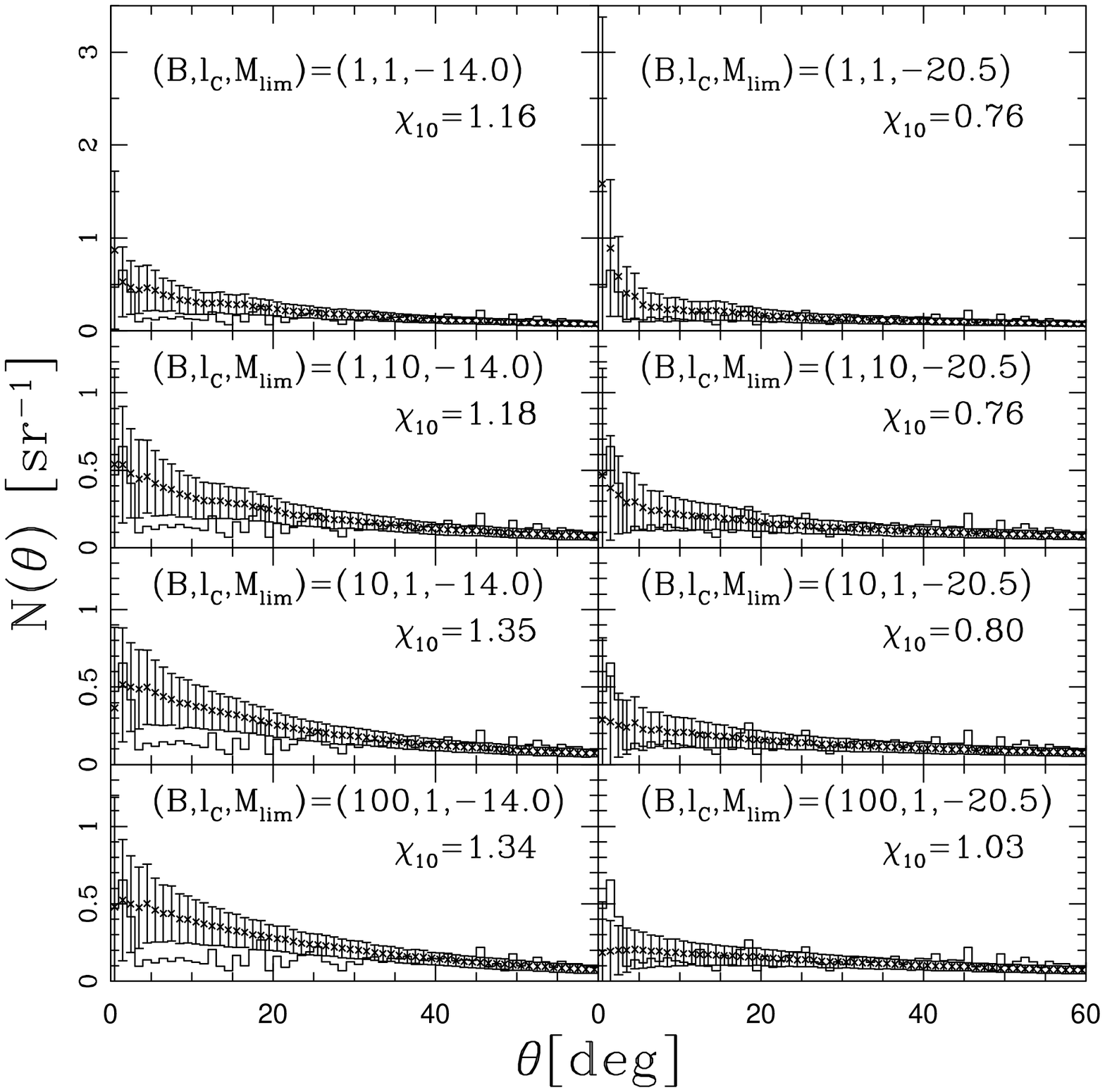}}}}
\figcaption{
The two point correlation function for number weighted sources
in the case of $M_{\rm{lim}} = -14.0$ (left panels), $-20.5$ (right panels)
and $(B,l_{\rm c})= (1,1), (1,10), (10,1), (100,1)$ in descending order.
The number of simulated events is set to be 57 with energies of
$(10^{19.6}-10^{20.3})$ eV.
The histograms represent the AGASA data in this energy range.
$\chi_{10}$ is also shown.
\label{fig13}}
\vspace{0.5cm}

At larger angles $(\sim 5^{\circ})$, there are relatively strong correlation
for $M_{\rm{lim}} = -14.0$, which is not consistent with the observed isotropic
distribution, whereas arrival distributions for $M_{\rm{lim}} = -20.5$
are sufficiently isotropic.
Among the cases of $M_{\rm{lim}} = -20.5$,
noticeable feature is coexistence of a strong correlation at small angles
and sufficiently isotropic distribution at larger scale
for $(B,l_{\rm c})=(1,1)$.
This is a consequence of the sharp peak of the UHECR angular image
in the case of $(B,l_{\rm c})=(1,1)$ (see Figure~\ref{fig5}).
This strong correlation at small angles is reduced or eliminated
for longer correlation length or stronger EGMF.
Due to small number of observed events, the statistics of
the two point correlation function is limited.
For this reason, $\chi_{ 10 }$ for $(B,l_{\rm c})=(1,1)$
is equal to that for $(B,l_{\rm c})=(1,10)$.
However, if the future experiments like the Pierre Auger
array \citep*{capelle98} confirm the small scale clustering
of UHECR arrival directions with more event number,
the case of $(B,l_{\rm c})=(1,1)$ will be favored
with better significance.

Let us summarize the results of the three quantities
(energy spectrum, harmonic amplitude, two point correlation function).
First, the energy spectrum for $B=1$ nG provide
better fit to the AGASA data than that for strong EGMF
in terms of $\chi_{ES}$.
Among this case, the extension of the energy spectrum
is also reproduced for $M_{\rm {lim}} \sim -14.0$,
although there is significant large scale anisotropy
contrary to the AGASA data in this case.
In the case of $M_{\rm {lim}} \sim -20.0$, the GZK cutoff
is predicted.

Second, the amplitude of the first harmonics is consistent
with the AGASA data at $M_{\rm {lim}} \sim -20.5$
irrespective of $(B,l_{\rm c})$.
The amplitude of the second harmonics is not consistent
with the AGASA for any $(B,l_{\rm c},M_{\rm {lim}})$.

Finally, the two point correlation function for
$(B,l_{\rm c},M_{\rm {lim}})=(1,1,-20.5)$ indicate
coexistence of a strong correlation at small angles $(\sim 1^{\circ})$
and sufficiently isotropic distribution at larger scale.
In sum, the source model of
$(B,l_{\rm c},M_{\rm {lim}})=(1,1,-20.5)$ seems to reproduce
the observation better than another parameter sets.
The amplitude of the second harmonics and the extension
of the energy spectrum can not be explained in this source model.

\subsection{Dependence on the Number of UHECR Sources}\label{number}

Up to present, we have not explained the large scale isotropy
quantified by the amplitude of the second harmonics
and the small scale clustering except for $(B,l_{\rm c})=(1,1)$,
as well as the extension of the cosmic ray spectrum above $10^{20}$ eV.
Since we assumed that all the galaxies more luminous than a given
limiting magnitude contribute to the cosmic ray flux,
the number of sources is maximal and the fluctuations around
the assumed non-isotropic distribution is minimal,
which make the anisotropy most visible.
Furthermore, each source contribute at most one event in this case,
it is difficult to obtain
clusters which are likely to reflect the point-like sources.
Given this situation, we select some galaxies from our ORS sample, and
investigate dependence of the results on the number of selected galaxies.
It is also expected that we can know the effect of fluctuation of
source number within the GZK sphere on the energy spectrum
above $10^{20}$ eV.
As an example, this selection corresponds to restricting the
sources of UHECRs to host galaxies of GRBs, which contribute
to the currently observed cosmic ray flux.

We assumed that contributions to arrival distribution from sources
outside 8000 km s$^{-1}$ from us,
within which the ORS is sampling the universe, is completely isotropic one.
However, since we select only some galaxies in the following,
point-like nature of sources has to be taken into account.
Thus, we distribute galaxies homogeneously in this region,
and including them, we select some galaxies from our ORS sample.
We restrict ourselves to the limiting magnitude of $M_{\rm{lim}}=-20.5$
and number weighted sources,
which produce the most isotropic arrival distribution of UHECRs.
It is noted that the number density of galaxies more luminous than
$M_{\rm{lim}}=-20.5$ in our ORS sample is
$\sim 10^{-4.3}$ Mpc$^{-3}$.

In Figure~\ref{fig14}, we show the amplitude of the first harmonics
as a function of the number fraction (NF) of selected sources to all
the ORS galaxies more luminous than $M_{\rm{lim}}=-20.5$
for all the $(B,l_{\rm c})$.
For each number fraction, we plot the average over all trial of
source selection and realization from the simulated probability distribution
with two error bars.
The smaller is the statistical error due to the finite number
of observed events, while the larger is both the statistical error
and the cosmic variance, that is, variation between different
selections of sources from our ORS sample.
In order to obtain the average and variance, we dial the simulated sets
of events 100 times from probability distribution
predicted by a specific source distribution, and
the sets of sources 30 and 10 times from our ORS sample for
the number fraction of $< 0.1$ and $> 0.1$
respectively in the case of $B=1,10$ nG, but 30 times for
all the number fractions in $B=100$ nG.
We note that the number of sources, which contributes to the cosmic ray
flux, for each set of $(B,l_{\rm c})$ differ from the others, even
for the same number fraction, because of
the different range of UHECRs (See Figure~\ref{fig4}).
For example, there is no source within 100 Mpc for the number fraction
of $< 10^{-2.5}$ (for the number density of $< 10^{-7}$ Mpc$^{-3}$)
in the case of $B=100$ nG.

\vspace{0.5cm}
\centerline{{\vbox{\epsfxsize=7.5cm\epsfbox{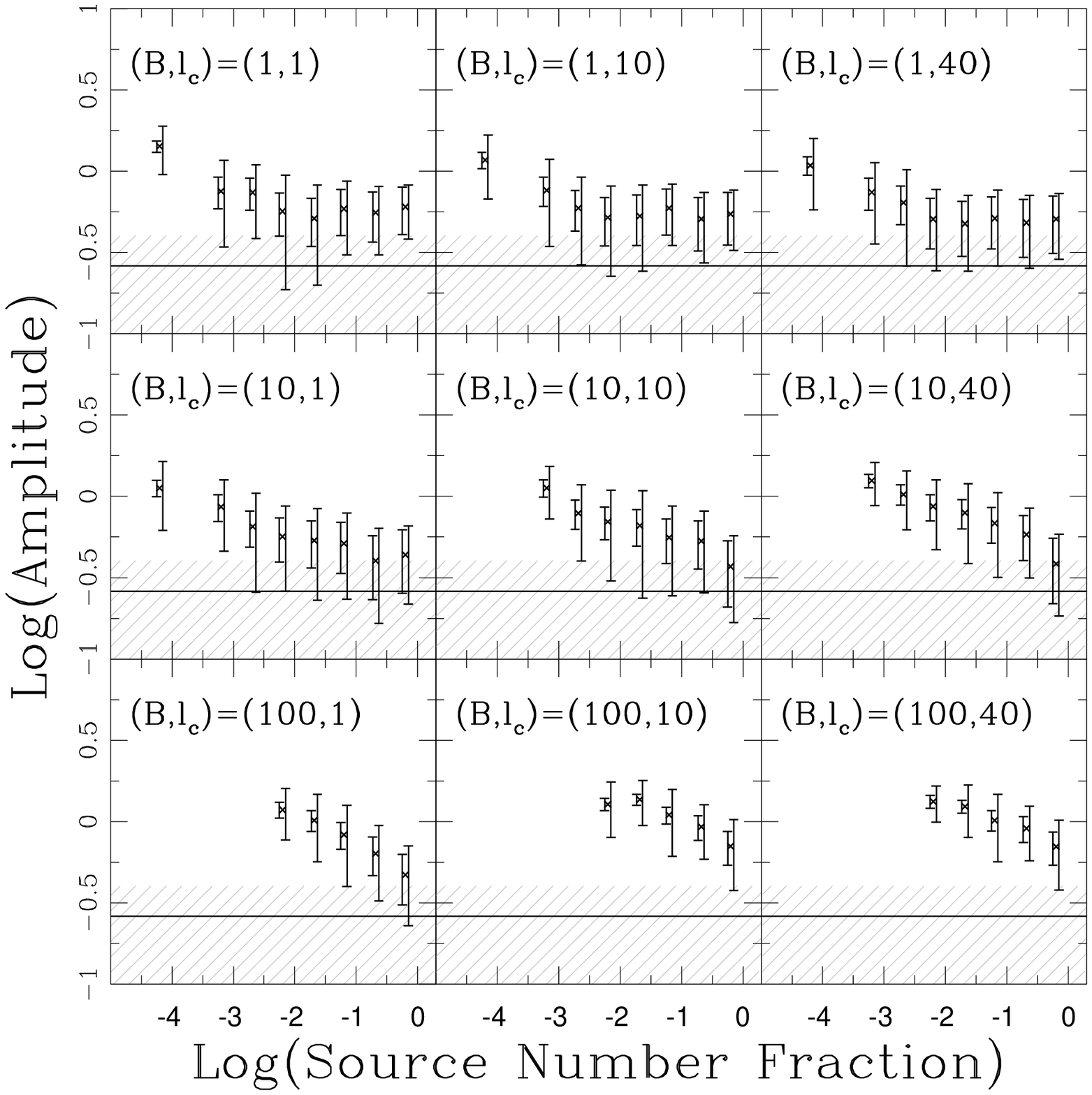}}}}
\figcaption{
Amplitude of the first harmonics as a function of the number fraction
of UHECR sources relative to the ORS galaxies luminous
than $M_{\rm{lim}}=-20.5$.
UHECR sources are selected from the ORS sample of $M_{\rm{lim}}<-20.5$.
The number of simulated events is set to be 57 in the energy range of
$(10^{19.6}-10^{20.3})$.
The points indicate the realization average, and the left and right
error bars represent the statistical and total error, respectively.
The solid line represent the AGASA data in the corresponding energy range.
The shaded region is expected from the statistical fluctuation
of isotropic source distribution with the chance probability
larger than $10 \%$.
\label{fig14}}
\vspace{0.5cm}

As a general trend, anisotropy increases with decreasing
the number fraction, because the arrival directions of UHECRs
clumps to that of sources which is small in number.
For large number fraction $(\sim 1)$, the amplitudes approach to the values
in the case that all galaxies more luminous than $M_{\rm{lim}}=-20.5$
contribute to the observed cosmic ray flux, as they should.
As clearly visible in Figure~\ref{fig4},
UHECRs can propagate the longest distance for $(B,l_{\rm c})=(1,1)$,
in which the deflection angle, and then the effective path length,
is smallest among for another parameters.
Accordingly, distant sources make a substantial contribution
to the arrival distribution of UHECRs, unlike in the case of stronger EGMF.
This is reflected in the decrease of the amplitude around
the number fraction of $10^{-1.7}$ for $(B,l_{\rm c})=(1,1)$.
Except for $B=100$ nG, arrival distribution is sufficiently
isotropic compared to isotropic source distribution
for almost all the number fractions.

The amplitude of the second harmonics is also shown in Figure~\ref{fig15}.
Dependence on the number fraction is roughly same as that of
the first harmonics.
For $B=1$ and $10$ nG, the predicted amplitude is
sufficiently isotropic for the number fraction
of $(10^{-3.0}-10^{-1.5})$.
On the other hand, it appears to be inconsistent with the
observation for any number fraction in the case of $B=100$ nG.

Figure~\ref{fig16} shows $\chi_{10}$ of the two point
correlation function as a function of the number fraction.
Increase of deviations between models and data with decreasing
the number fraction is attributed to the clump of UHECRs
at directions of their sources, similarly to the case of the harmonic analysis.
Especially, this dependence is quite noticeable for $(B,l_{\rm c})=(1,1)$,
where UHECRs hardly deflect so that correlation at small angle
is too strong.
At larger number fraction ($\ge 10^{-2.0}$),
it can be seen that the model of
$(B,l_{\rm c})=(1,1)$ provide relatively better fit to the observation
than another models.

\vspace{0.5cm}
\centerline{{\vbox{\epsfxsize=7.5cm\epsfbox{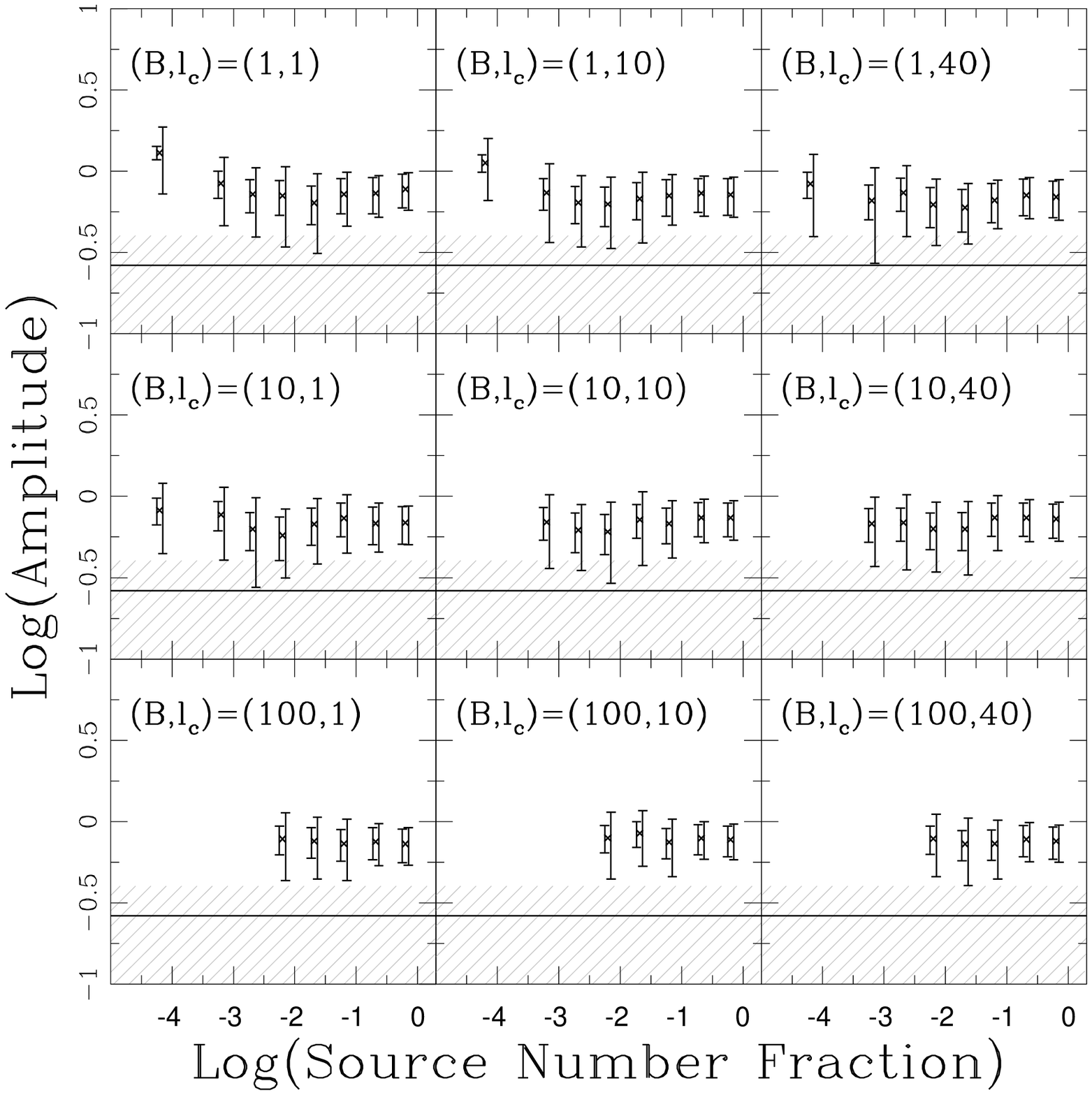}}}}
\figcaption{
Same as Figure~\ref{fig14}, but amplitude of the second harmonics.
\label{fig15}}
\vspace{0.5cm}

In Figure~\ref{fig17}, we show the two point correlation
function predicted by a specific source scenario for
$(B,l_{\rm c})=(1,1)$, $(1,10)$ and $(10,1)$
with the number fraction of $\sim 10^{-1.7}$.
$\chi_{10}$ is also shown.
Each source distribution is selected so that it predicts the smallest
value of $\chi_{10}$ for respective set of $(B,l_{\rm c})$.
Only for $(B,l_{\rm c})=(1,1)$, another condition is also imposed
that the harmonic amplitude predicted by the source model
is consistent with the AGASA observation within $1 \sigma$ level.
The correlation of events at small angle scale and sufficient isotropic
distribution at larger scale are well reproduced for $(B,l_{\rm c})=(1,1)$.
There is no significant correlation for $(B,l_{\rm c})=(1,10)$ and $(10,1)$,
because of the larger deflection angle of UHECRs.
In the cases of strong EGMF or longer correlation length,
it appears to be difficult to reproduce the strong anisotropy
only at the smallest angle scale, even for the case of
minimal value of $\chi_{10}$.

\vspace{0.5cm}
\centerline{{\vbox{\epsfxsize=7.5cm\epsfbox{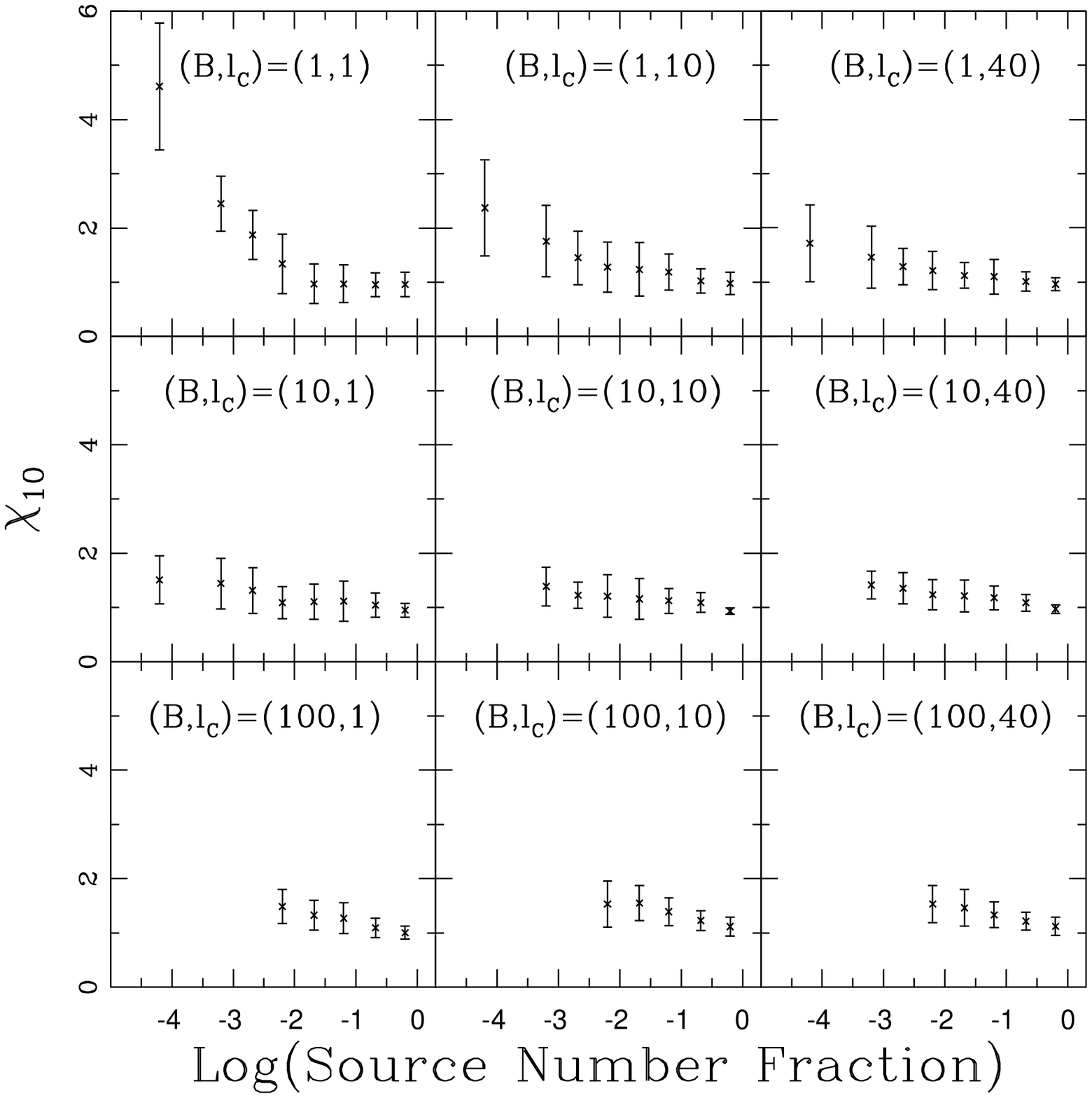}}}}
\figcaption{
$\chi_{10}$ as a function of the number fraction
of UHECR sources relative to the ORS galaxies luminous
than $M_{\rm{lim}}=-20.5$.
UHECR sources are selected from the ORS sample of $M_{\rm{lim}}<-20.5$.
The errorbars represent cosmic variance due to different realizations
of the source selection.
\label{fig16}}
\vspace{0.5cm}

\vspace{0.5cm}
\centerline{{\vbox{\epsfxsize=7.5cm\epsfbox{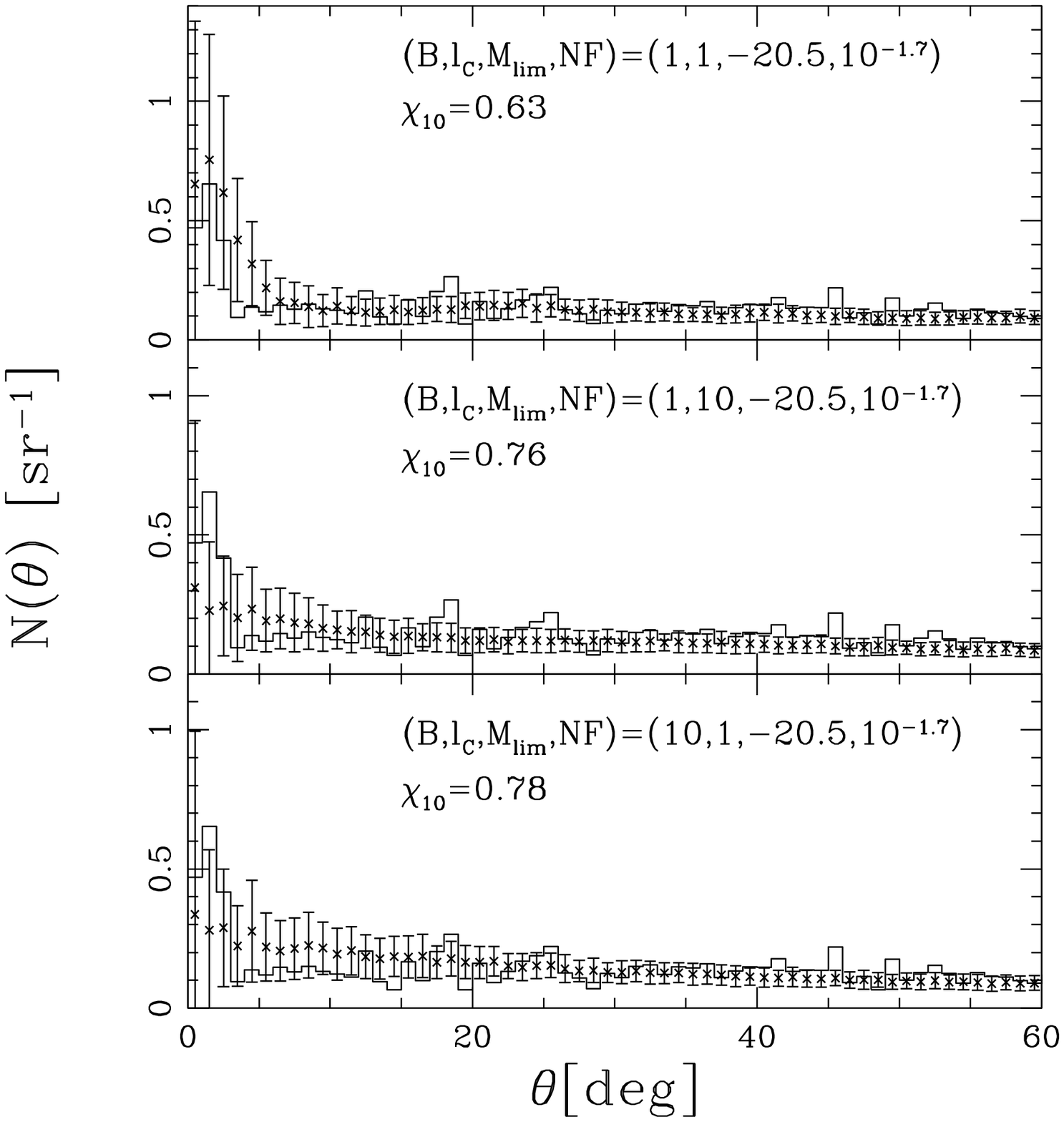}}}}
\figcaption{
The two point correlation function predicted by a specific
source scenario in the case that the number fraction (NF)
$\sim 10^{-1.7}$
of the ORS galaxies more luminous than $M_{\rm {lim}}=-20.5$
is selected as UHECR sources.
The number of simulated events is set to be 57 with energies of
$(10^{19.6}-10^{20.3})$ eV.
The histograms represent the AGASA data in this energy range.
$\chi_{10}$ is also shown.
\label{fig17}}
\vspace{0.5cm}

Considering the calculated results of the first and second harmonics
and the small scale anisotropy all together,
the model of $(B,l_{\rm c})=(1,1)$ and the number fraction of $\sim 10^{-1.7}$
(the source number density of $\sim 10^{-6}$ Mpc$^{-3}$)
seems to reproduce the observations better than another parameter sets.
However, due to the small number of observed events, the statistics for
the two point correlation function is limited.
For this reason, $\chi_{10}$ does
not differ very much from each other.
The future experiments like the Pierre Auger array \citep*{capelle98}
will decrease statistical uncertainty and provide more strong constraints
to the model predictions.

\vspace{0.5cm}
\centerline{{\vbox{\epsfxsize=7.5cm\epsfbox{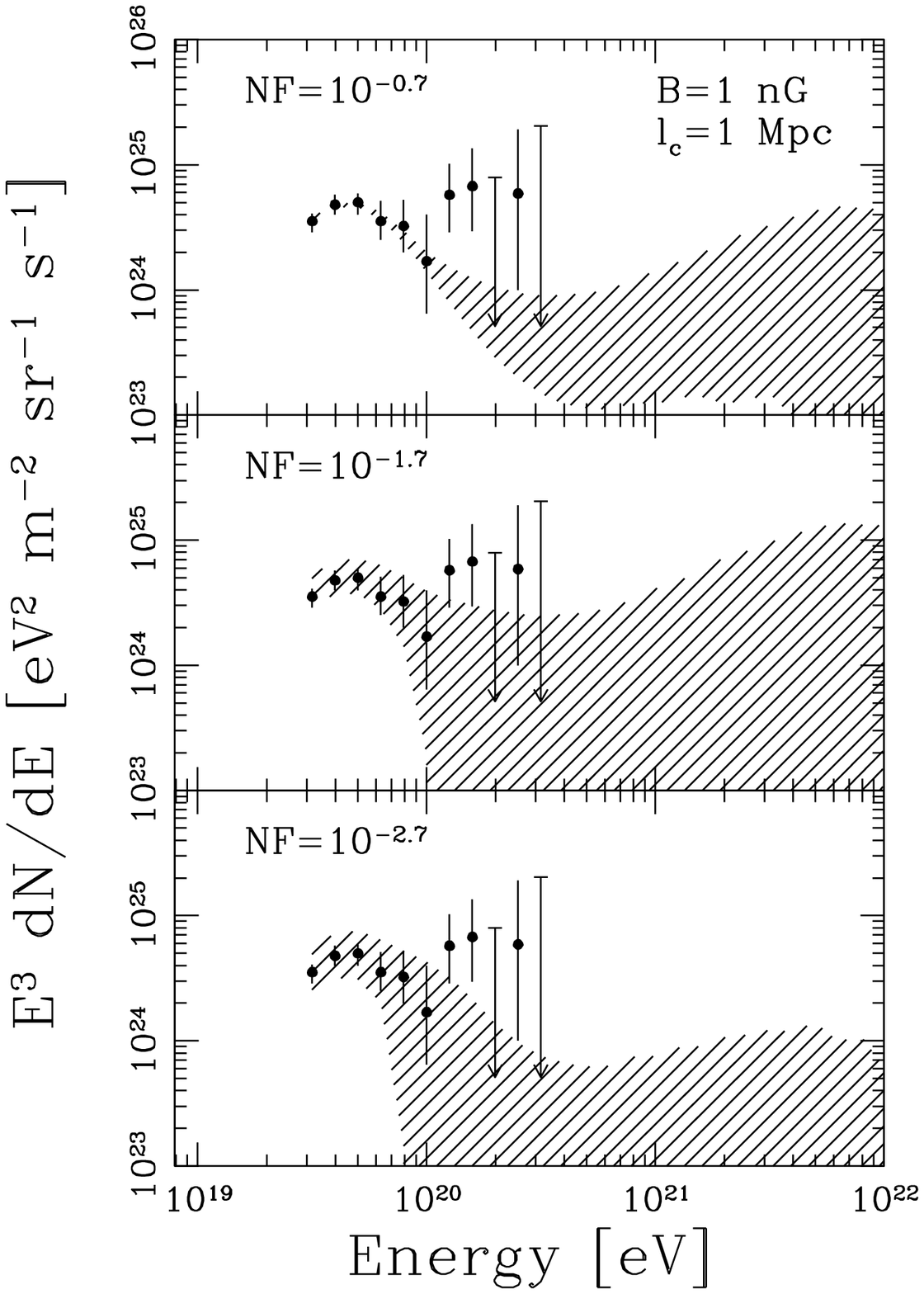}}}}
\figcaption{
The energy spectra predicted by sources selected from the ORS
galaxies more luminous than $M_{\rm {lim}}=-20.5$
in the case of $(B,l_{\rm c})=(1,1)$.
NF represent the number fraction of selected UHECR sources
to all the ORS galaxies $(M_{\rm {lim}}<-20.5)$.
They are normalized so as to minimize $\chi_{ES}$.
The shaded regions represent 1 $\sigma$ error due to the source
selection from our ORS sample.
\label{fig18}}
\vspace{0.5cm}

Next we discuss dependence of the energy spectrum on the number fraction.
In Figure~\ref{fig18}, we show the predicted energy spectra in the case
of $(B,l_{\rm c})=(1,1)$ for the number fraction $10^{-0.7}$,
$10^{-1.7}$, $10^{-2.7}$ in descending order.
They are normalized so as to minimize $\chi_{ES}$.
The shaded regions represent 1 $\sigma$ error due to the source
selection from our ORS galaxies more luminous than $M_{\rm {lim}}=-20.5$.
As is evident from this figure, decreasing the number fraction
increases fluctuation of the predicted energy spectrum,
especially above $10^{20}$ eV.
For the number fraction of $10^{-2.7}$ (the source number density of
$\sim 10^{-7}$ Mpc$^{-3}$), since there are no source
within the GZK sphere, the GZK cutoff is clearly visible.
The predicted energy spectrum is roughly consistent with
the AGASA observation within 1 $\sigma$ level
in the case of the number fraction of $10^{-1.7}$, where
the observed isotropy and clusters are also reproduced well.

However, this result should be interpreted with care.
Since the source number density is $\sim 10^{-6}$ Mpc$^{-3}$
in the case of the number fraction of $10^{-1.7}$, mean number of
sources is $\sim 0.5$ within the GZK sphere.
Therefore, the AGASA 8 events above $10^{20}$ eV, which do not constitute
the clustered events with each other, must originate from
at most a few sources.
We found a source model which explain the extension of the energy spectrum.
In this source model, there are two sources within the region covered by the
ORS ($\sim 107$ Mpc), one at $\sim 20$ Mpc, the other at $\sim 105$ Mpc.
In Figure~\ref{fig19}, we show the energy spectrum predicted by this
source model.
The dotted line represent contribution from the nearest source.
The predicted energy spectrum provide good fit to the observed one
including the extension beyond $10^{20.0}$ eV.
However, almost all the events above $10^{20.0}$ are generated
by the nearest source as evident from Figure~\ref{fig19}.
These events are strongly concentrated in a single center,
which is inconsistent with the AGASA 8 events.
It seems to be difficult to reproduce the AGASA observation
above $10^{19.6}$ eV (including above $10^{20}$ eV) by a single
scenario of UHECR origin, even for
$(B,l_{\rm c},M_{\rm {lim}},NF)=(1,1,-20.5,10^{-1.7})$.

\vspace{0.5cm}
\centerline{{\vbox{\epsfxsize=7.5cm\epsfbox{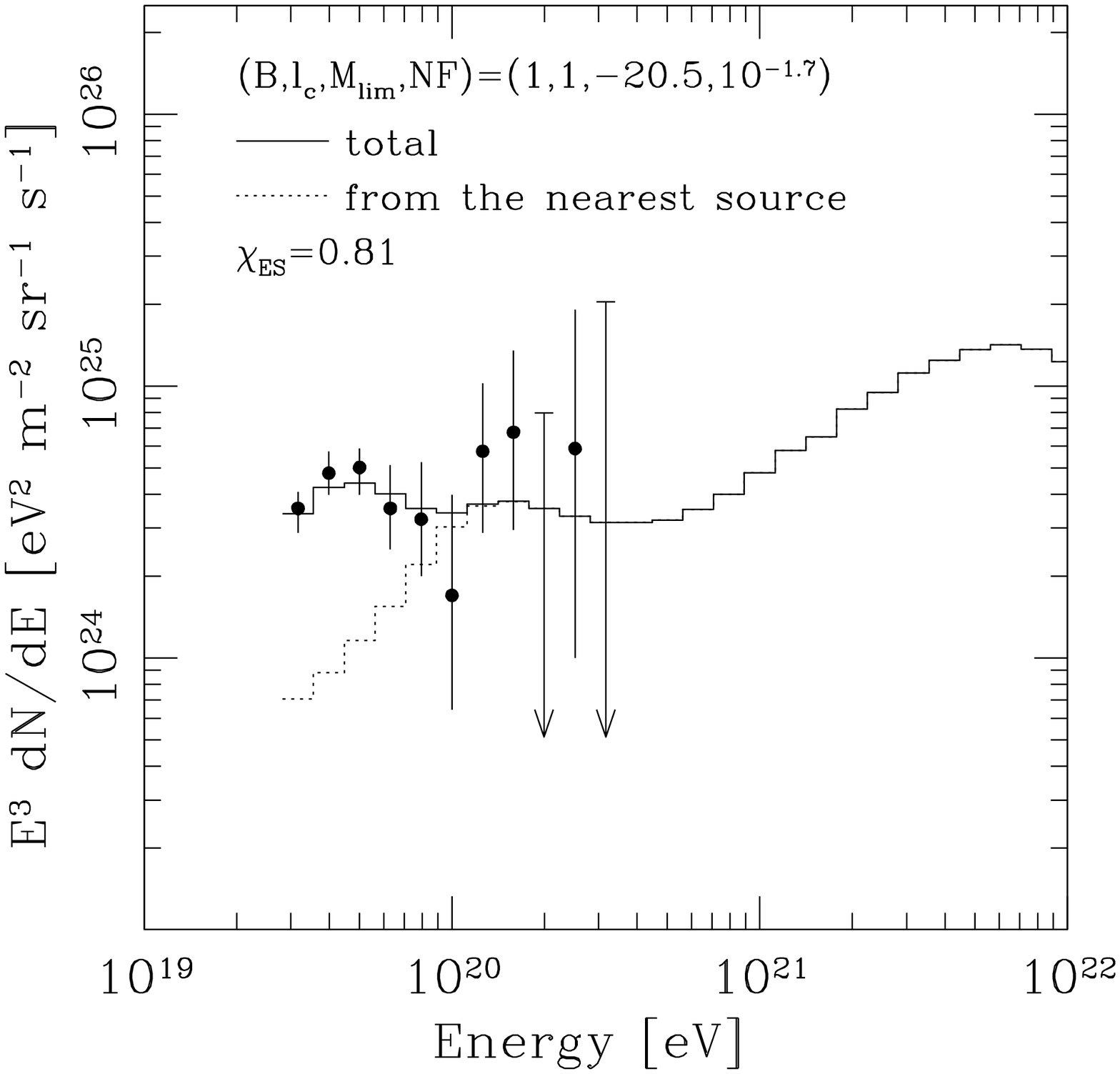}}}}
\figcaption{
The energy spectrum predicted by a specific
source scenario in the case that the number fraction (NF)
$\sim 10^{-1.7}$ of the ORS galaxies more luminous than
$M_{\rm {lim}}=-20.5$ is selected as UHECR sources.
The dotted line represent contribution from the nearest source
(at $\sim 20$ Mpc) in the source model.
The energy spectrum is normalized so as to
minimize $\chi_{ES}$.
The energy spectrum observed by the AGASA is also shown.
\label{fig19}}
\vspace{0.5cm}

\vspace{0.5cm}
\centerline{{\vbox{\epsfxsize=7.5cm\epsfbox{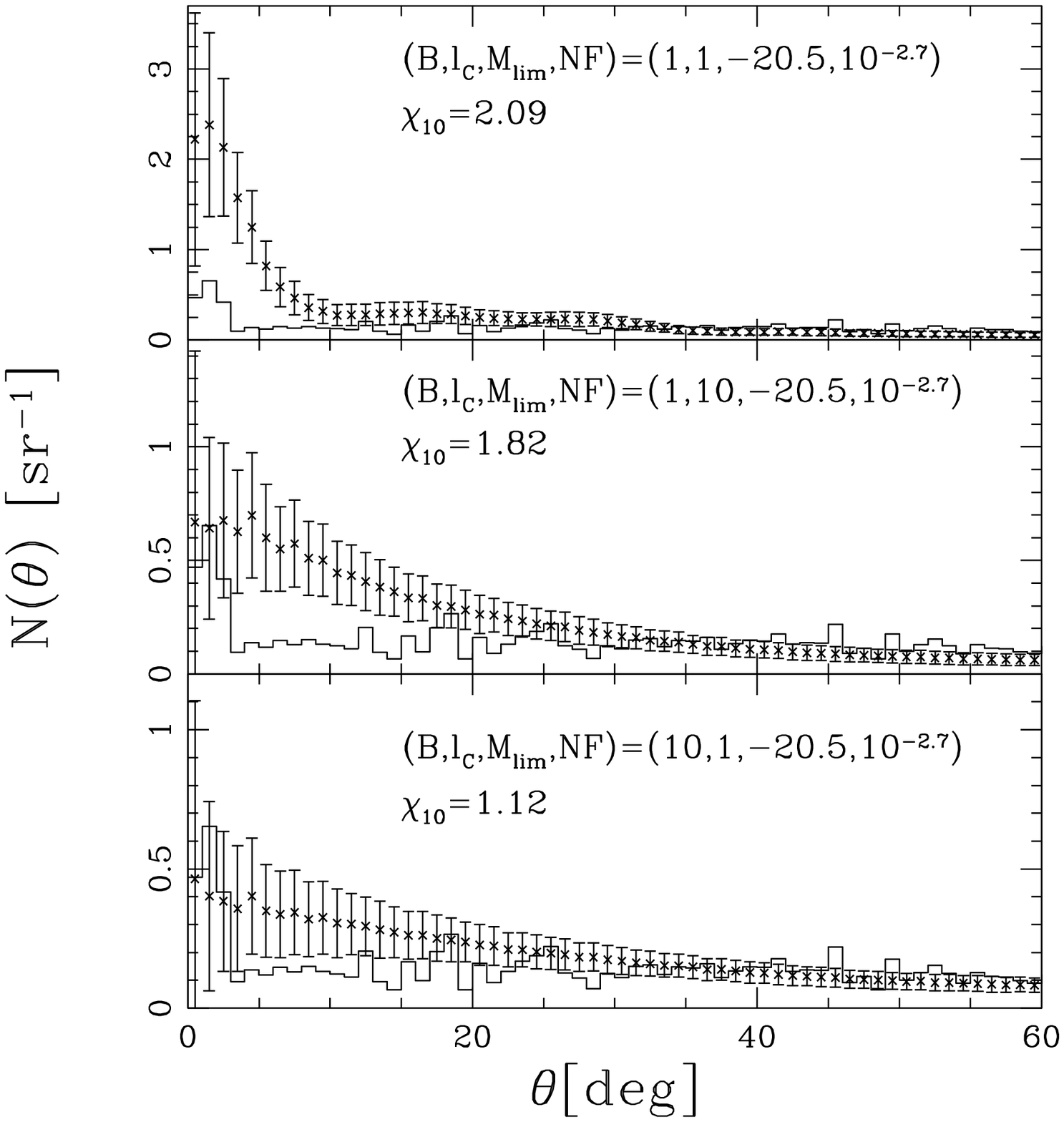}}}}
\figcaption{
Same as Figure 17, but for the
number fraction of $\sim 10^{-2.7}$.
\label{fig20}}
\vspace{0.5cm}

Recently, the HiRes Collaboration \citep*{wilkinson99}
reported the cosmic ray flux from $2 \times 10^{17}$ eV to over $10^{20}$
eV including the GZK cutoff \citep*{abu02}.
In our numerical simulations, the sharp GZK cutoff of the energy
spectrum is predicted for the number fraction of
$\sim 10^{-2.7}$ as shown above.
The small scale correlation of events in this case is too strong
for $(B,l_{\rm c})=(1,1)$ as is evident from Figure~\ref{fig20},
where the two point correlation function is shown in the same way
as Figure~\ref{fig17}, but with the number fraction of $\sim 10^{-2.7}$.
This strong correlation is reduced for $(B,l_{\rm c})=(1,10)$ and $(10,1)$,
but still inconsistent with the AGASA data.

\vspace{0.5cm}
\centerline{{\vbox{\epsscale{1.0}\plotone{fig21.eps}}}}
\figcaption{
Arrival directions of UHECRs with energies of $(10^{19.6}-10^{20.3})$ eV
in the equatorial coordinate predicted by the source model
of $(B,l_{\rm c},M_{\rm {lim}},NF)=(1,1,-20.5,10^{-1.7})$ of Figure 17.
For a comparison, the AGASA 57 events in this energy range are also shown.
The number of simulated events is also set to be 57.
The arrival directions are restricted in the range
$-10^{\circ} \le \delta \le 80^{\circ}$ in order to compare our results
with the AGASA data.
$r_{\rm m}$ represent the m-th harmonic amplitude.
\label{fig21}}
\vspace{0.5cm}

Finally we show in Figure~\ref{fig21} realizations of 57 arrival
directions of UHECRs with energies of $(10^{19.6}-10^{20.3})$ eV,
predicted by the source model of
$(B,l_{\rm c},M_{\rm {lim}},NF)=(1,1,-20.5,10^{-1.7})$
of Figure~\ref{fig17}.
The AGASA 57 events in this energy range are also shown.
The arrival directions are restricted in the range
$-10^{\circ} \le \delta \le 80^{\circ}$ in order to compare our results
with the AGASA data.
The arrival distribution of UHECRs
seems to be isotropic on a large scale with clusters of events
meaning the small scale anisotropy.
The amplitudes of the first and second harmonics are consistent
with the AGASA, but the extension of the spectrum is not.
(Mean number of simulated events above $10^{20}$ eV is only 1.)
In this example, C6 doublet at
$(\alpha, \delta) \sim (210^{\circ}, 35^{\circ})$
is well reproduced.
Also, high intensity regions
around $(\alpha, \delta) \sim (180^{\circ}, 10^{\circ})$
mentioned in Section 3.1. are eliminated under the favor
of the source selection.

\section{DISCUSSION} \label{discussion}

\subsection{Constraint on UHECR source model}\label{constraint}

We presented the results of numerical simulations on the
propagation of UHE protons with energies of $(10^{19.5}-10^{22})$ eV
injected by a discrete distribution of galaxies using the ORS data sample,
which is properly corrected for the selection effect
and absence of galaxies in the zone of avoidance ($|b|<20^{\circ}$).
We adopted strength of the EGMF to be not only 1 nG but also 10,100 nG,
in order to explore the possibility of reproducing the
extension of the energy spectrum above $10^{20}$ eV.
We calculated the three observable quantities, cosmic ray spectrum,
harmonic amplitude, and two point correlation function.
With these quantities, we made comparisons between the model
predictions and the existing AGASA data.

At first, we explored the source model consistent with the current
observations, as a function of the limiting magnitudes of galaxies.
The energy spectrum for the strong EGMF shows the extension
beyond $10^{20}$ eV.
However, it can not be well fitted with the observed
energy spectrum, especially at $E<10^{20}$ eV where the current data
have adequate statistics.
On the other hand, the energy spectrum for $B=1$ nG provide
good fit with the observation below $E=10^{20}$ eV, and
statistical significance of deviation $\chi_{ES}$ is smaller than that
for the strong EGMF.
Among this case, the extension of the energy spectrum
is also reproduced for $M_{\rm {lim}} \sim -14.0$,
although there is significant large scale anisotropy
contrary to the AGASA data in this case.
In the case of $M_{\rm {lim}} \sim -20.0$, the GZK cutoff
is predicted.

We found that the angular image became to be isotropic and
showed good correlation with AGASA events
as restricting sources to more luminous galaxies.
As a consequence, galaxies more luminous than $-20.5$ mag produce
the angular image of UHECR which is most isotropic roughly irrespective
of strength and correlation length of the EGMF.
However, it is not isotropic enough to be consistent with the observed
amplitude of the second harmonics, even for $M_{\rm{lim}}=-20.5$.
We further found that the small scale anisotropy is well reproduced
in the case of $(B,l_{\rm c})=(1,1)$ because of the small deflection
angle of UHECRs.

Next, in order to obtain sufficiently isotropic arrival
distribution of UHECRs and small scale clustering even for
the case of other than $(B,l_{\rm c})=(1,1)$,
we randomly selected sources from our ORS sample,
and investigate the dependence of the results on their number.
We found that the arrival distribution
became to be sufficiently isotropic in the case that the number fraction
$(10^{-3.0}-10^{-1.5})$ of the ORS galaxies more luminous than
$M_{\rm{lim}}=-20.5$ is selected as UHECR sources
for $B=1$, $10$ nG.
Among this case, the small scale clustering can also be reproduced
for $(B,l_{\rm c},M_{\rm {lim}},NF)=(1,1,-20.5,10^{-1.7})$
better than the cases of another parameter sets.
We furthermore found that the extension of the energy spectrum
is explained in this case within roughly 1 $\sigma$ error
due to source selection from our ORS sample
because of the large fluctuation above $10^{20}$ eV.

In sum, we showed that the three observable quantities including
the GZK cutoff of the energy spectrum can be reproduced
in the case that the number fraction $\sim 10^{-1.7}$ of the ORS
galaxies more luminous than $-20.5$ mag is selected as UHECR sources.
In terms of the source number density, this constraint
corresponds to $\sim 10^{-6}$ Mpc$^{-3}$.

\subsection{Implication for sources of UHECR above $10^{20}$ eV}
\label{implication}

As mentioned in the previous subsection, we showed that
the three observable quantities including
the GZK cutoff of the energy spectrum can be reproduced
in the case that the number fraction $\sim 10^{-1.7}$ of the ORS
galaxies more luminous than $-20.5$ mag is selected as UHECR sources.
However, we should interpret this result with care.
Since the AGASA 8 events above $10^{20}$ eV do not constitute
the clustered events with each other, the number of their sources
within the GZK sphere must be much larger than the event number (8).
On the other hand, the number of sources within the GZK sphere is at most
only 1 in the case of the number fraction of $10^{-1.7}$
(the source number density of $10^{-6}$ Mpc$^{-3}$).
It is suspected that cosmic rays above $10^{20}$ eV (hereafter EHECRs;
extremely-high energy cosmic rays) are strongly concentrated
in a single center, even if the extension of the energy spectrum is
explained.
Therefore, large fraction of EHECRs observed by the AGASA
might originate from sources other than ones of UHECR
with energies of $(10^{19.6}-10^{20.0})$ eV.
The number density of EHECR source
must be much larger than $10^{-6}$ Mpc$^{-3}$.

We can put forward two candidates as such EHECR production site.
It is suggested that iron ions from the surfaces
of young strongly magnetized neutron stars may be accelerated to
$10^{20}$ eV through relativistic MHD winds \citep*{blasi00}.
However, it seems to be difficult to obtain sufficiently isotropic arrival
directions of EHECRs from sources localized in the Galactic plane
by deflection due to the Galactic magnetic field \citep*{oneill01}.

The decay of some supermassive particles, which could be produced from TDs,
or be certain MSRPs, are also considered as probable EHECRs origin.
Decay rate of the supermassive particles required to explain the
observed EHECR flux are estimated as $\sim 10^{35}$ Mpc$^{-3}$ yr$^{-1}$
\citep*{bhattacharjee00}.
Multiplying by the current EHECR observing time ($\sim 10$ yr),
we obtain the number density of $\sim 10^{36}$ Mpc$^{-3}$,
which is sufficiently large to reproduce not clustered AGASA 8 EHECRs.
As for the large scale anisotropy,
distribution of TDs in the universe would be homogeneous.
MSRPs are expected to be clustered in galactic halos, and
thus EHECR flux will be dominated by contribution from our own Galactic Halo.
Accordingly, arrival distribution would be isotropic
enough to be consistent with the current AGASA observation in both cases.
Furthermore, these scenarios are not constrained by contribution
to the cosmic ray flux at lower energy which must be smaller
than observed one, since hard injection spectrum is generally predicted.
For the reasons stated above, the top-down scenarios may be favored
to explain large fraction of the 8 EHECRs observed by the AGASA.

There is a problem if all the EHECRs observed by the AGASA
originate from sources other than those of UHECRs.
If we consider that sources of EHECRs and UHECRs 
precisely differ from each
other, there would be no clustered event set which includes both
EHECRs and UHECRs, because clustered events are expected
to reflect a single point source. 
However, 2 EHECRs out of 8 observed by the AGASA constitute the clustered
events with UHECRs with energies of $(10^{19.6}-10^{20})$ eV
(C1 and C3 in Hayashida et al.(2000)).
We have to interpret that the two clusters happen to be
produced by chance.
Chance probability of more than two accidental clusterings within
$2.5^{\circ}$ between 49 UHECRs and 8 EHECRs is quite low $(\sim 4 \%)$.
The AGASA observation seems to deviate from our prediction with $\sim
2 \sigma$ confidence level.

However, about 1 EHECR out of 8 may originate in the bottom-up scenarios.
Indeed, 57 cosmic rays above $10^{19.6}$ eV predicted by
the source model of Figure~\ref{fig21} include an EHECR.
In this case, since the source of this single event will be located
within the GZK sphere, it is likely that this event constitutes a clustered
event set with UHECRs.
Chance probability of more than one accidental clustering rise up
to $\sim 28 \%$, which is high enough for us to
interpret one of the two doublets to be produced merely by chance.
Thus, our scenario is consistent with the current AGASA observation
within 1 $\sigma$ level.

The event correlation between EHECRs and UHECRs would provide an
important test of our conclusion with sufficient amount of data from
future experiments.
For example, when the event number of UHECRs above $10^{19.6}$
increases to 10 times as many as that of the current AGASA observation
(57), the number of EHECRs which lie within $2.5^{\circ}$ from UHECRs
by chance becomes $18.6 \pm 4.3 \,\,(1 \sigma)$.
If future experiments observe event number of such EHECRs within this
range, our scenario will be supported.
Here we note that the source number within the GZK sphere is at most
1 in the case of NF=$10^{-1.7}$.
Therefore the number of event clustering between an EHECR and an UHECR
which can be explained by bottom-up scenarios is at most 1.

Another test of our conclusion is distinguishing the chemical composition
of primary cosmic rays which initiate air showers in the earth atmosphere.
In the bottom-up scenarios, protons or nuclei are primary particles.
On the other hand, photons are considered as primary
in the top-down scenarios \citep*{bhattacharjee00}.
Information on the chemical composition is mainly provided by 
the depth of shower maximum for fluorescence observation of the air shower.
There are a number of projects utilizing the fluorescence
technique, such as HiRes \citep*{wilkinson99},
South and North Auger \citep*{capelle98},
OWL \citep*{owl00} and EUSO \citep*{euso92}.
These experiment would provide information on the chemical composition
and a test of our conclusion.

\subsection{Comparison with the HiRes cosmic ray spectrum}
\label{hires}

The HiRes Collaboration \citep*{wilkinson99}
reported the cosmic ray flux from $2 \times 10^{17}$ eV to over $10^{20}$
eV including the GZK cutoff \citep*{abu02}.
As discussed above, we showed that the three observable quantities
except for EHECR events observed by the AGASA can be reproduced
in the case that the number fraction $\sim 10^{-1.7}$ of the ORS
galaxies more luminous than $-20.5$ mag is selected as UHECR sources.
In Figure~\ref{fig22}, we show the predicted energy spectrum
fitted to the data measured by Hires-I detector
in the case of $(B,l_{\rm c},M_{\rm {lim}},NF)=(1,1,-20.5,10^{-1.7})$.
As easily seen, the fit is good as compared with the middle panel
of Figure~\ref{fig18}.
If the flux measured by the HiRes Collaboration
is correct and observational features
about the arrival distribution are same as the AGASA,
our source model can explain both the
arrival distribution of UHECRs and the flux at the same time.
When the large-area detectors like the Pierre Auger array \citep*{capelle98}
start operating, the statistical error of the observed energy spectrum
will decrease.
This allows us to draw a conclusion about
the origin of EHECRs or whether they exist or not.

\vspace{0.5cm}
\centerline{{\vbox{\epsfxsize=7.5cm\epsfbox{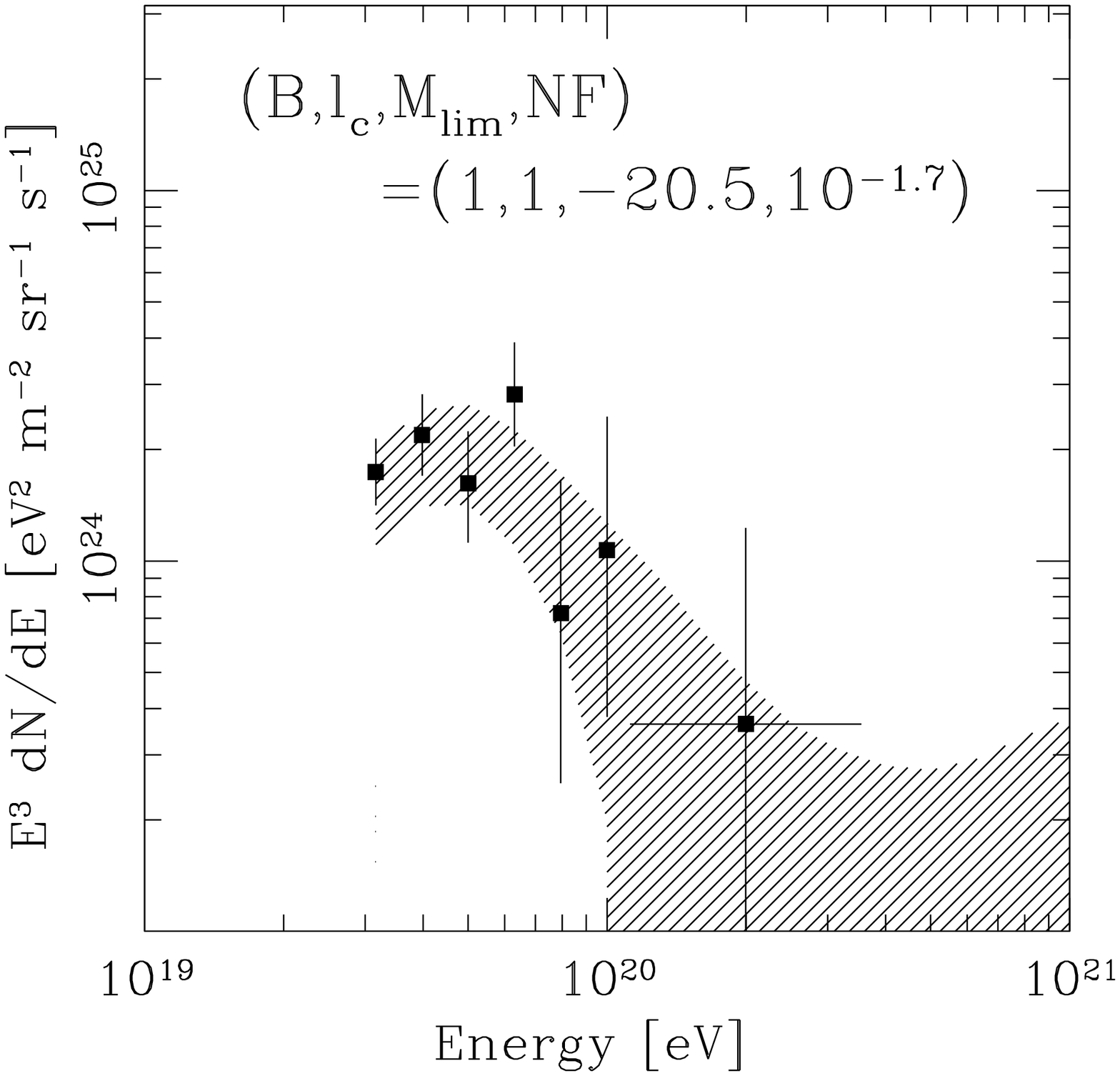}}}}
\figcaption{
The energy spectra predicted by sources selected from the ORS
galaxies more luminous than $M_{\rm {lim}}=-20.5$
in the case of $(B,l_{\rm c},NF)=(1,1,-20.5,10^{-1.7})$.
NF represent the number fraction of selected UHECR sources
to all the ORS galaxies $(M_{\rm {lim}}<-20.5)$.
They are fitted to the data of HiRes-I detector (squares and errorbars).
The shaded regions represent 1 $\sigma$ error due to the source
selection from our ORS sample.
\label{fig22}}
\vspace{0.5cm}

\subsection{Implication for sources of UHECR below $10^{20}$ eV}
\label{implication2}

\begin{table*}
\caption{Number density of possible UHECR production site
\label{tbl:num}}

\begin{center}
\begin{tabular}{cccc}
\tableline\tableline\noalign{\smallskip}
 Object & Density & Ratio & ref \\
 & [Mpc$^{-3}$] & to our result & \\
 & & ($10^{-6}$ Mpc$^{-3}$) & \\
\noalign{\smallskip}
\tableline\noalign{\smallskip}
AGN                & $1 \times 10^{-4}$ &  100.0  & \cite{loveday92} \\
FR-II radio galaxy & $3 \times 10^{-8}$ &   0.03  & \cite{woltjer90} \\
BL Lac             & $3 \times 10^{-7}$ &   0.3   & \cite{woltjer90} \\
GRB                & $1 \times 10^{-4}$ &  100.0  & \cite{mao98}     \\
dead quasar        & $5 \times 10^{-4}$ &  500.0  & \cite{boldt99}   \\
colliding galaxy   & $7 \times 10^{-7}$ &   0.7   & \cite{smi02}     \\
\noalign{\smallskip}
\tableline
\noalign{\smallskip}

\end{tabular}
\end{center}
\end{table*}

In this subsection we discuss implications of our results, obtained by
comparison of the model predictions with the existing AGASA data,
for the origin of UHECRs with energies of $(10^{19.6}-10^{20})$ eV.
The number densities of possible UHECR production sites
and ratios to our result ($\sim 10^{-6}$ Mpc$^{-3}$)
are tabulated in Table~\ref{tbl:num}.

Blast waves in AGN jet could accelerate particle to a few tens of EeV
\citep*{halzen97} and similarly for AGN cores.
However, due to the interaction of UHECRs with very high radiation fields
in and around the central engine of an AGN, these maxima are unlikely
to be achieved.
This is consistent with our result, because the number density of AGN
are much larger than our results, and thus, arrival distribution
may be inconsistent with the AGASA if all the AGNs contribute to 
the cosmic ray flux above $10^{19.6}$ eV.
Among AGNs, the most promising acceleration sites for UHECRs are the
so-called hot-spots of FR-II radio galaxies
\citep*{biermann97,norman95}.
However, their number density seems to be too small,
as compared with our result.
It is unlikely that FR-II radio galaxies are responsible for
all the AGASA 49 events with energies of $(10^{19.6}-10^{20})$ eV.
Another possible UHECRs production sites are BL Lacertae objects
(BL Lacs) \citep*{tinyakov01}.
Their number are larger than FR-II radio galaxies,
but still insufficient to reproduce all the 49 events.

GRB can be shown to in principle accelerate protons up to
$10^{20}$ eV \citep*{waxman00}.
The effective density of GRB is determined by the GRB rate
$(\sim 2 \times 10^{-10}$ h$^3$ Mpc$^{-3}$ yr$^{-1}$ \citep*{mao98})
and the typical time delay $\tau$ of UHECRs, which depends
on the propagation distance from sources to the earth.
Since we are now considering UHECRs with energies of $(10^{19.6}-10^{20})$ eV,
we have to include contribution from sources outside the
GZK sphere when calculating the typical time delay.
We calculated the mean time delay from our numerical data assuming
homogeneous GRBs distribution, and obtained $\tau \sim 10^6$ yr.
This gives the GRB effective density $\sim 10^{-4}$ Mpc$^{-3}$.
From our result, if about 1 $\%$ of GRBs produce UHECRs, arrival
distribution predicted by GRBs would be consistent with the AGASA.

The compact dynamo model has been proposed as a natural mechanism
for accelerating cosmic rays in dead quasars
\citep*{boldt99,levinson00,boldt00}.
Only 0.2 $\%$ of dead quasars are required to contribute to
the comic ray flux above $10^{19.6}$ eV.

Large fraction of luminous infrared galaxies (LIRGs)
are found to be interacting systems,
suggesting that collisions and merging processes are responsible for
triggering the huge IR light emission \citep*{sanders88}.
In such systems,
favorable environments for accelerating cosmic rays to $10^{20}$ eV
are provided by amplified magnetic fields on the scale of
tens kpc resulting from gravitational compression, as well as high
relative velocities of galaxies and/or superwinds from multiple
supernovae explosions \citep*{cesarsky92}.
According to Smialkowski et al.(2002),
we take the number density of such colliding galaxies
as that of galaxies with huge far infrared luminosity
$L_{\rm {FIR}}>10^{11}L_{\odot}$.
The number density is roughly equal to our result.
If all the galaxies with $L_{\rm {FIR}}>10^{11}L_{\odot}$ are responsible
for UHECRs, predicted arrival distribution would be consistent with
the AGASA.
Indeed,  Smialkowski et al.(2002) showed that arrival distribution of UHECRs
predicted by LIRGs is sufficiently isotropic such that
statistical tests they used for large scale
anisotropy was not conclusive for distinguishing between isotropic
and LIRGs source distribution.
Furthermore, they showed that the probability of occurring the clustered
events are more than 10 times higher for LIRGs source distribution
than that for isotopic distribution.
Their results well agree with ours.

\subsection{Consistency with a statistical analysis of event cluster}
\label{consistency}

Our result is supported by a statistical analysis of clustering of
UHECRs developed by Dubovsky et al.(2000),
which allow an estimate of the minimum number density of sources.
They applied their arguments to the 14 events above $10^{20}$
eV observed by several experiments.
They obtained the minimum number density of sources
$\sim 6 \times 10^{-3}$ Mpc$^{-3}$.
Since we are now considering UHECRs below $10^{20}$ eV,
we have to apply the arguments given in their paper
to the cosmic ray events below $10^{20}$ eV.

Let us apply the arguments to the 49 events
with energies of $(10^{19.6}-10^{20.0})$ eV observed by the AGASA.
According to them, the number density must be larger than
\begin{eqnarray}
h_* = \frac{1}{4 R^3} \cdot
\frac{N_{ \rm{tot} }^3}{N_{ \rm{cl} }^2},
\label{minnum}
\end{eqnarray}
where $N_{ \rm{tot} }$, $N_{ \rm{cl} }$, $R$ are the total number of events,
the number of events in clusters, e-folding length of cosmic ray
luminosity from a single source respectively.
The value of $R$ can be obtained as $\sim 700$ Mpc from Figure~\ref{fig4}.
Taking $N_{ \rm{tot} }=49$ and $N_{ \rm{cl} }=9$,
Eq.(~\ref{minnum}) gives $h_* \sim 10^{-6}$ Mpc$^{-3}$
which is consistent with our result $\sim 10^{-6}$ Mpc$^{-3}$.

\subsection{Strong EGMF in the LSC}
\label{effect}

Throughout the paper, we assumed that the statistical properties of
the EGMF are uniform.
However, simple analytical arguments based on magnetic flux freezing,
and large scale structure simulations passively including the magnetic
field \citep*{kulsrud97} demonstrate that the magnetic field is most
likely as structured as are the baryons.
The EGMF as strong as $\sim 1 \mu$G in sheets and filaments of large
scale galaxy distribution, such as in the LSC, are compatible with existing
upper limits on Faraday rotation \citep*{ryu98,blasi99}.
It is suspected that the arrival distribution of UHECRs depends on the
fields in the immediate environment of the observer.

However, numerical simulations of UHECR propagation in inhomogeneous
EGMF over cosmological distances are highly time-consuming.
With the assumption of homogeneous EGMF, we have performed numerical
simulation of spherically symmetric propagation.
Provided that the EGMF is inhomogeneous, the propagation of UHECRs
from a single point source has no longer spherical symmetry.
As a result, we have to specify the earth position in the universe.
We also have to choose the detector (earth) size so small enough for
us to accurately calculate the arrival directions.
In this case, the number fraction of UHECRs arriving at the earth to
injected ones is extremely small.
This requires the number of particle to be propagated several orders
of magnitudes higher than that used in this study, which takes
enormous CPU time.

Although the strong EGMF in the LSC is compatible with the Faraday
ratation measures, the cosmic ray observations do not appear to
support such local EGMF.
Indeed, we showed in this study that small scale clustering can be
well reproduced in the case that UHECRs propagate along nearly
straight lines.
If the local strong EGMF affects the arrival directions of UHECRs,
small scale clustering observed by the AGASA will not be obtained.

This difficulty can be seen from the studies by another authors.
There are several groups who study the propagation of UHECR in such
strong EGMF in the LSC \citep*{sigl99,lemoine99,isola02,sigl02}.
They assume simplified source distributions which represent the LSC.
The strong EGMF of $\sim 1 \mu$G leads to substantial deflections of UHECRs,
which are better for explaining the observed isotropic arrival
distribution of UHECRs.
However, consistency of small scale anisotropy and also large scale
isotropy predicted by their scenarios with the AGASA observation
is somewhat worse than that predicted by our scenario.
The strong EGMF in the LSC seems to be unfavored by the current cosmic
ray observation, although it is not ruled out by the Faraday rotation
measurements.

Of course, since the event number of UHECRs is currently very small,
we can not absolutely rule out such strong EGMF in the LSC.
Fulture experiments would increase the significance level of statement
about whether such strong EGMF affects the arrival directions of
UHECRs or not.

\section{CONCLUSION} \label{conclusion}

In this section, we present our conclusions obtained
from comprehensive study on the bottom-up scenarios.
For the origin of EHECRs,
we conclude that large fraction of the AGASA 8 events
above $10^{20}$ eV might
originate in the topdown scenarios, or that the cosmic ray flux
measured by the HiRes experiment might be better.
We also discussed the origin of UHECRs below $10^{20.0}$ eV
through comparisons between the number density of astrophysical
source candidates and our result ($\sim 10^{-6}$ Mpc$^{-3}$).
At present, we can not conclude which source candidates
are responsible for UHECRs below $10^{20.0}$ eV.
However, our result for the source number density
($\sim 10^{-6}$ Mpc$^{-3}$) will be useful
when searching for the ultimate UHECR source.

Throughout the paper, we set constraints to the source models
so that their predictions reproduce the current AGASA observation
within 1 $\sigma$ error of the statistical fluctuation
due to finite number of the observed events.
However, if we relax the constraints to within 2 $\sigma$
error, the energy spectrum (Figure~\ref{fig9} and ~\ref{fig10})
and the amplitude of the second harmonics (Figure~\ref{fig12})
can be reproduced in the case of
$(B,M_{\rm {lim}}) \sim (1,-20.5)$ without the source selection.
In this case, since the source number density is
$\sim 5.4 \times 10^{-5}$ Mpc$^{-3}$, there are $\sim 30$ sources
within the GZK sphere.
The AGASA events above $10^{20}$ eV, which do not constitute
the clustered events with each other, may also be reproduced,
contrary to the case of the source number density of $\sim 10^{-6}$ Mpc$^{-3}$.
Thus, we can not absolutely rule out the bottom-up scenarios
as the origin of EHECRs.

The statistics for the quantities which we used is limited by the
small number of observed events, which make difficult
to draw a decisive conclusion on the origin of UHECRs.
However, the present working or development of large-aperture
new detectors, such as HiRes \citep*{wilkinson99}
and South and North Auger \citep*{capelle98}
will considerably decrease the statistical uncertainties.
There are furthermore plans for space based air shower detectors such as
OWL \citep*{owl00} and EUSO \citep*{euso92} which would detect about
1000 events per year above $10^{20}$ eV \citep*{bhattacharjee00}.
With the method to constrain the source model and their number density
developed in this paper, these experiments will reveal the
origin of UHECRs with better significance in the very near future.

\acknowledgments
We would like to thank Dr. T. Stanev for kindly giving us the data
of interaction length and inelasticity of UHE protons due
to the photopion production with photons of the cosmic microwave background.
This research was supported in part by Giants-in-Aid for Scientific
Research provided by the Ministry of Education, Science and Culture
of Japan through Research Grant No.S14102004 and No.S14079202.




\begin{thebibliography}{}
\bibitem[Abu-Zayyad et al. 2002]{abu02} Abu-Zayyad T. et al.
(The HiRes Collaboration) 2002, astro-ph/0208243

\bibitem[Benson and Linsley 1992]{euso92} Benson R., Linsley J.
1992, \aap, 7, 161

\bibitem[Berezinsky et al. 2002]{berezinsky02} Berezinsky V., Gazizov A.Z.,
Grigorieva S.I. 2002, hep-ph/0204357

\bibitem[Bhattacharjee and Sigl 2000]{bhattacharjee00} Bhattacharjee P.,
Sigl G. 2000, Phys. Rep. 327, 109

\bibitem[Biermann 1997]{biermann97} Biermann P.L. 1997, Nucl. Part. Phys.,
23, 1

\bibitem[Blanton et al. 2001]{blanton01} Blanton M., Blasi P.,
Olinto A.V. 2001, Astropart. Phys., 15, 275

\bibitem[Blasi et al. 1999]{blasi99} Blasi P., Burles S.,
Olinto A.V. 1999, \apj, 514, L79

\bibitem[Blasi et al. 2000]{blasi00} Blasi P., Epstein R.I.,
Olinto A.V. 2000, \apj, 533, L123

\bibitem[Blasi and Olinto 1998]{blasi98} Blasi P., Olinto V. 1998, Phys. Rev.
D, 59, 023001

\bibitem[Boldt and Ghosh 1999]{boldt99} Boldt E., Ghosh P. 1999,
Mon. Not. R. Astron. Soc. 307, 491

\bibitem[Boldt and Loewenstein 2000]{boldt00} Boldt E., Loewenstein M. 2000,
Mon. Not. R. Astron. Soc. 316, L29

\bibitem[Capelle et al. 1998]{capelle98} Capelle K.S., Cronin J.W.,
Parente G., Zas E. 1998, APh, 8, 321

\bibitem[Cesarsky 1992]{cesarsky92} Cesarsky C.J. 1992, Nucl. Phys. B
(Proc. Suppl.), 28, 51

\bibitem[Chodorowski et al. 1992]{chodorowski92} Chodorowski M.J.,
Zdziarske A.A., Sikora M. 1992, \apj, 400, 181 

\bibitem[Cline and Stecker 2000]{owl00} Cline D.B., Stecker F.W.
OWL/AirWatch science white paper, astro-ph/0003459

\bibitem[Dubovsky et al. 2000]{dubovsky00} Dubovsky S.L., Tinyakov P.G.,
Tkachev I.I. 2000, Phys. Rev. Lett., 85, 1154


\bibitem[Greisen 1966]{greisen66} Greisen K. 1966, Phys. Rev. Lett., 16, 748 

\bibitem[Halzen and Zas 1997]{halzen97} Halzen F., Zas E. 1997, \apj,
488, 669

\bibitem[Hayashida et al. 1999]{hayashida99} Hayashida N., et al. 1999
APh, 10, 303

\bibitem[Hayashida et al. 2000]{hayashida00} Hayashida N., et al. 2000,
astro-ph/0008102

\bibitem[Hillas 1984]{hillas84} Hillas A.M. 1984, Ann. Rev. Astron.
Astrophys., 22, 425

\bibitem[Ide et al. 2001]{ide01} Ide Y., Nagataki S., Tsubaki S.,
Yoshiguchi H., Sato K. 2001, Publ. Astron. Soc. Japan, 53, 1153

\bibitem[Isola et al. 2002]{isola02} Isola C., Sigl G. 2002,
astro-ph/0203273

\bibitem[Kronberg 1994]{kron94} Kronberg P.P. 1994, Rep. Prog. Phys.
57, 325

\bibitem[Kulsrud et al. 1997]{kulsrud97} Kulsrud R.M., Cen R.,
Ostriker J.P., Ryu D. 1997, \apj, 480, 481

\bibitem[Lemoine et al. 1997]{lemoine97} Lemoine M., Sigl G., Olinto A. V.,
Schramm D.N. 1997, \apj, 486, L115

\bibitem[Lemoine et al. 1999]{lemoine99} Lemoine M., Sigl G., Biermann P.
1999, astro-ph/9903124

\bibitem[Levinson 2000]{levinson00} Levinson A. 2000, Phys. Lev. Lett.,
85, 912

\bibitem[Loveday et al. 1992]{loveday92} Loveday J., et al.
1992, \apj, 390, 338

\bibitem[Mao and Mo 1998]{mao98} Mao S., Mo H.J. 1998,
\aap, 339, L1

\bibitem[Mucke et al. 2000]{sophia00} Mucke A., Engel R., Rachen J.P.,
Protheroe R.J., Stanev T. 2000, Comput. Phys. Commun. 124, 290

\bibitem[Norman et al. 1995]{norman95} Norman C.A., Melrose D.B.,
Achterberg A. 1995, \apj, 454, 60

\bibitem[O'Neill et al. 2001]{oneill01} O'Neill S., Olinto A.,
Blasi P. 2001, astro-ph/0108401


\bibitem[Ryu et al. 1998]{ryu98} Ryu D., Kang H., Biermann P.L., 1998,
\aap, 335, 19

\bibitem[Sanders et al. 1988]{sanders88} Sanders D.B., Soifer B.T.,
Elias J.H., Madore B.F., Matthews K., Neugebauer G.,
Scoville N.Z. 1988, \apj, 325, 74

\bibitem[Santiago et al. 1995]{santiago95} Santiago B.X., Strauss M.A.,
Lahav O., Davis M., Dressler A., Huchra J.P. 1995, \apj, 446, 457

\bibitem[Santiago et al. 1996]{santiago96} Santiago B.X., Strauss M.A.,
Lahav O., Davis M., Dressler A., Huchra J.P. 1996, \apj, 461, 38

\bibitem[Selvon 2000]{selvon00} Selvon A.L. 2000, astro-ph/0009444

\bibitem[Sigl et al. 1999]{sigl99} Sigl G., Lemoine M., Biermann P.
1999, Astropart. Phys., 10, 141

\bibitem[Sigl 2002]{sigl02} Sigl G. 2002, astro-ph/0210049

\bibitem[Smialkowski et al. 2002]{smi02} Smialkowski A., Giller M.,
Michalak W. 2002, astro-ph/0203337

\bibitem[Stanev et al. 2000]{stanev00} Stanev T., Engel R., Mucke A.,
Protheroe R.J., Rachen J.P. 2000, Phys. Rev. D, 62, 093005

\bibitem[Takeda et al. 1998]{takeda98} Takeda M., et al. 1998, Phys.
Rev. Lett., 81, 1163

\bibitem[Takeda et al. 1999]{takeda99} Takeda M., et al. 1999, \apj, 522, 225

\bibitem[Tinyakov and Tkachev 2001]{tinyakov01} Tinyakov P.G., Tkachev I.I.
2001, JETP Lett., 74, 445

\bibitem[Vietri 1995]{vietri95} Vietri M. 1995, \apj, 453, 883


\bibitem[Waxman 1995]{waxman95} Waxman E. 1995, Phys. Rev. Lett., 75, 386

\bibitem[Waxman 2000]{waxman00} Waxman E. 2000, Nucl. Phys. Proc. Suppl.,
87, 345

\bibitem[Wilkinson et al. 1999]{wilkinson99} Wilkinson C.R., et al. 1999,
APh,  12, 121

\bibitem[Woltjer 1990]{woltjer90} Woltjer L. 1990,
$\it{Active}$ $\it{Galactic}$ $\it{Nuclei}$,
Blanford R.D., Netzer H., Woltjer L. (Springer-Verlag, Berlin) p.1

\bibitem[Yoshida and Teshima 1993]{yoshida93} Yoshida S., Teshima M.
1993, Prog. Theor. Phys. 89, 833

\bibitem[Yoshiguchi et al. 2002]{yoshiguchi02} Yoshiguchi H., Nagataki
S., Sato K., Ohama N., Okamura S. 2002, astro-ph/0212061

\bibitem[Zatsepin and Kuz'min 1966]{zatsepin66} Zatsepin G.T., Kuz'min V.A.
1966, JETP Lett., 4, 78
\end{thebibliography}
\end{document}